\documentclass[accepted]{uai2021} 

\usepackage[american]{babel}
\usepackage{natbib} 
\usepackage{amsfonts,amsthm}
\usepackage{amsmath}
\usepackage{booktabs}
\usepackage{subcaption}
\usepackage{caption}
\usepackage{hyperref}
\usepackage{amssymb}
\usepackage{multirow}
\usepackage[switch]{lineno} 
\usepackage{color}
\usepackage{xcolor,colortbl}

\newcommand{\mat}{\mathbf}

\definecolor{Gray}{gray}{0.85}
\newcolumntype{a}{>{\columncolor{Gray}}c}
\newcolumntype{b}{>{\columncolor{white}}c}
\usepackage{dsfont}

\newtheorem*{theorem}{Theorem}
\usepackage{bm}
\newcommand\independent{\protect\mathpalette{\protect\independenT}{\perp}}
\def\independenT#1#2{\mathrel{\rlap{$#1#2$}\mkern2mu{#1#2}}}

\title{Automated Meta-Analysis: A Causal Learning Perspective}

\author[1,2]{\href{mailto:Lu Cheng <lcheng35@asu.edu>}{Lu~Cheng}{}} 
\author[1]{Dmitriy~A.~Katz-Rogozhnikov}
\author[1]{Kush~R.~Varshney}
\author[1]{Ioana~Baldini}
\affil[1]{%
    IBM Research -- Thomas J. Watson Research Center\\
    Yorktown Heights, NY, USA
}
\affil[2]{%
    Computer Science and Engineering\\
    Arizona State University\\
    AZ, USA
}

\begin{document}
\maketitle

\begin{abstract}
Meta-analysis is a systematic approach for understanding a phenomenon by analyzing the results of many  previously published experimental studies. It is central to deriving conclusions about the summary effect of treatments and interventions in medicine, poverty alleviation, and other applications with social impact. Unfortunately, meta-analysis involves great human effort, rendering a process that is extremely inefficient and vulnerable to human bias. To overcome these issues, we work toward automating meta-analysis with a focus on controlling for risks of bias. In particular, we first extract information from scientific publications written in natural language. From a novel causal learning perspective, we then propose to frame automated meta-analysis -- based on the input of the first step -- as a multiple-causal-inference problem where the summary effect is obtained through \textit{intervention}. Built upon existing efforts for automating the initial steps of meta-analysis, the proposed approach achieves the goal of automated meta-analysis and largely reduces the human effort involved. Evaluations on synthetic and semi-synthetic datasets show that this approach can yield promising results. 
\end{abstract}

\section{Introduction}
Meta-analysis --- a tool to amplify statistical power by aggregating weighted information from multiple similar studies \citep{moher2009preferred} --- is increasingly used in all fields of scientific research such as public health, psychology, and social science to understand a variety of socially impactful interventions. The goal of meta-analysis is to reveal common trends by combining a sufficient number of related studies, even if those individual studies contain sources of error \citep{xiong2018machine}. One of the major fields of meta-analysis is evidence-based medicine, where multiple similar randomized control trials (RCTs) are studied before treatments enter clinical practice. For instance, such a study could refer to questions such as: What is the therapeutic association between Vitamin C and breast cancer (negative effect, no effect, or positive effect)? or What is the relationship between the risk factor phosphodiesterase type 5 (PDE5) inhibitor and the cardiac morphology?

\begin{figure}
\centering
    \includegraphics[width=\columnwidth]{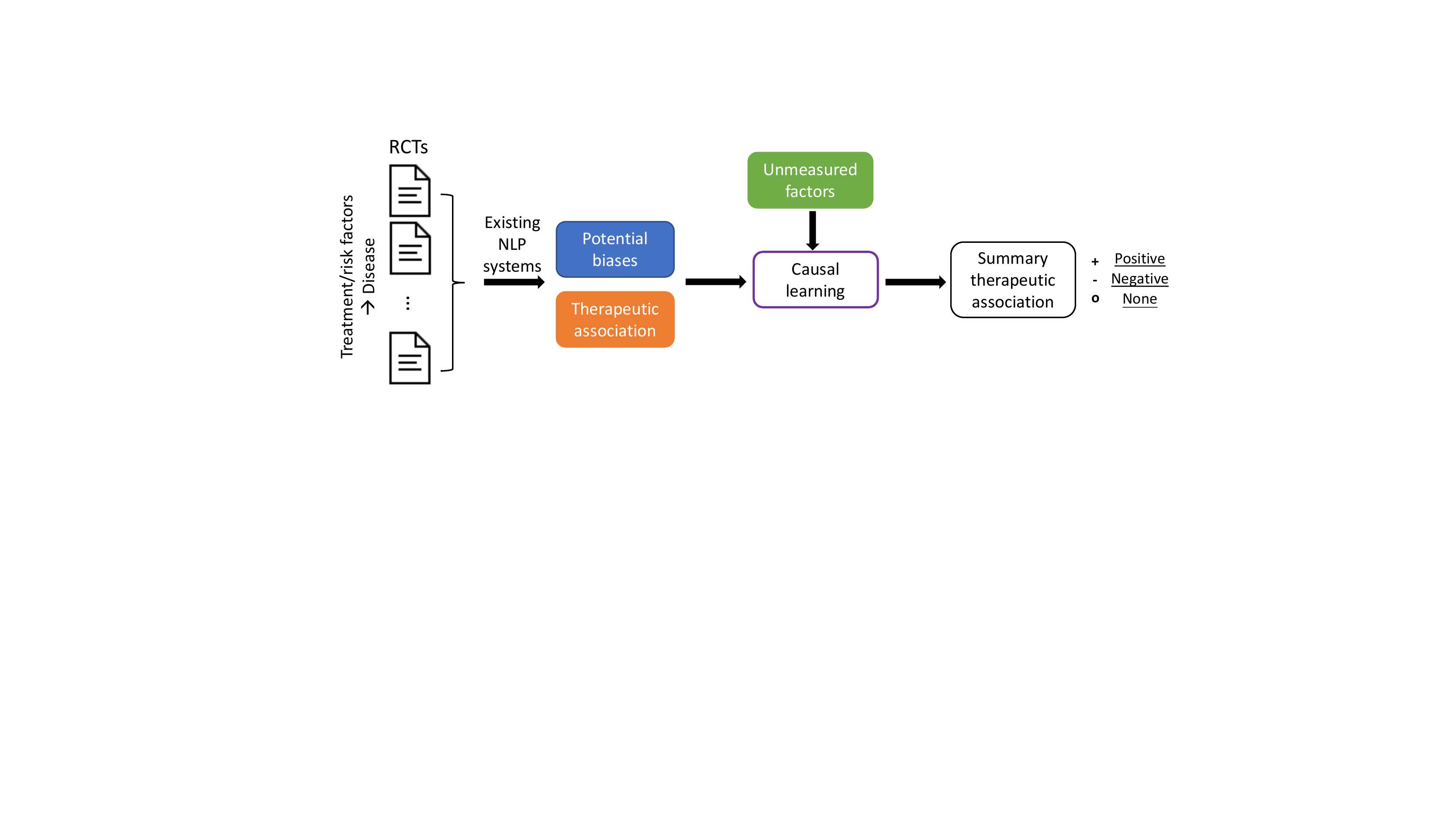}
  \caption{Block diagram for automating meta-analysis. `+' denotes positive effect, `-' negative effect, and `o' no effect.} 
  \label{problem}
\end{figure}

Despite its importance, meta-analysis suffers from limited coverage and practicality because the process is extremely time-consuming and labor-intensive. In a standard meta-analysis, a person does a comprehensive literature search, identifies relevant studies using inclusion/exclusion criteria, extracts data, and conducts statistical analysis. Undesired consequences can surface when applying manual meta-analysis such as unexplored topics and outdated RCTs \citep{michelson2014automating}. Manual meta-analysis is also prone to different sources of biases. For example, preconceived domain knowledge can influence the inclusion criteria designed by a meta-analyst, and individual studies with statistically non-significant results may never be made public, thereby skewing the meta-analytic samples \citep{carter2019deep}. An automated meta-analysis can lead to scalable procedures in which various potential biases are accounted for. Recent advances in natural language processing (NLP) \citep{xiong2018machine,marshall2016robotreviewer} have sought to automate the initial steps in meta-analysis by developing NLP systems to detect potential biases from scientific publications. 

Toward the overarching goal of automated meta-analysis, we need to take a step further to automatically infer the summary therapeutic association while controlling for these potential biases. Therefore, we work under a framework comprised of two major steps as depicted in Fig. \ref{problem}. However, we confront a primary challenge: these NLP systems do not extract the effect size and variance from the individual studies, which are the standard input to conventional meta-analysis models (e.g., random effect model \citep{borenstein2011introduction}). Instead, they output a \textit{discrete} therapeutic association and risks of biases detected in each RCT. Since conventional meta-analysis models are not applicable to the NLP outputs, we need a new approach for the second step to infer the summary therapeutic association based on the risks of bias and therapeutic association extracted from each RCT. Given that the inference might be further influenced by unmeasured factors, e.g., potential errors from the NLP system, it is necessary to account for these uncertainties in summarizing the estimation results.

To tackle these challenges, in this work, we provide a novel perspective of multiple causal inference that seeks to understand the causal relationship between risks of bias, i.e., the \textit{multiple treatments}, and the therapeutic associations we observe in different RCTs, i.e., the \textit{outcome}. From this causal perspective, we reduce the problem of inferring the summary therapeutic association to answering ``what the therapeutic association will be if no risk of biases are observed in the RCTs''; Moreover, drawing on causal inference theories, we can narrow down the potentially many unmeasured factors to those that simultaneously influence the risks of bias and therapeutic association, i.e., the \textit{hidden confounders}. Hidden confounders may cause spurious relationships between the treatment and outcome, rendering biased estimation of the summary therapeutic association. 

Our \textbf{contributions} are: we work on a practical problem of automating meta-analysis which has barely been studied in the literature. In contrast to conventional meta-analysis, we pay special attention to the influence of the risks of bias and the hidden confounders on estimating summary effects. We provide an innovative causality perspective for the problem and propose a novel causal learning module to infer the summary therapeutic association, significantly reducing the required human effort in conventional meta-analysis. We evaluate the proposed model on both synthetic and semi-synthetic data, the generating process of which is guided by real-world meta-analyses. We show that the proposed approach can improve the precision of the estimated summary therapeutic association. 
\section{Related Work}
\textbf{Meta-Analysis in Medical Research.} The need to make trustworthy clinical decision-making has fostered the momentum toward evidence-based medicine and meta-analysis \citep{sackett1996evidence}. There are two core elements of a standard meta-analysis: (1) effect size that reflects the magnitude of the treatment effect of an intervention, and (2) study weight that indicates the importance of the individual clinical trial \citep{borenstein2011introduction}. One aim of meta-analysis is to determine if an effect exists; the other aim is to determine if the effect is positive or negative. 

A major challenge of meta-analysis is the time-consuming and labor-intensive process. Consequently, there have been a few efforts toward developing automated meta-analysis \citep{michelson2014automating,xiong2018machine,carter2019deep}. For example, \citet{michelson2014automating} envisioned an automatic process for creating meta-analyses for any treatment and disease, and keeping them up-to-date automatically. It consists of three steps: paper extraction, paper clustering, and standard meta-analysis models. Later, \citet{xiong2018machine} developed an NLP model seeking to identify relevant publications.

Another major limitation of meta-analysis is the potential biases that have long been threats to the validity of meta-analytic results \citep{carter2019deep}. One such source of bias at the meta-analysis level is the inclusion and exclusion criteria of an RCT. At the level of individual studies, bias can be introduced by the loss of trials and subjects \citep{haidich2010meta}, publication bias (studies with significant, positive results are more likely to be published), and poor concealment of treatment allocation or no blinding in studies. All these biases can potentially influence the estimated treatment effects \citep{pildal2007impact}. 

The RobotReviewer of \citet{marshall2016robotreviewer} includes NLP techniques to estimate the therapeutic association and biases of an RCT from its text, which constitutes the first step before our proposed causal modeling. Our contributions herein complement the earlier work. In particular, we explore and demonstrate the inference of the summary therapeutic association \textit{in the presence of potential biases and unmeasured factors from the outputs of NLP techniques such as RobotReviewer} by leveraging a unique multiple causal inference perspective that enables the combination of an advanced NLP system with causal learning.

\noindent\textbf{Multiple Causal Inference.} 
One seeks to identify the effects of \textit{multiple} treatment on the outcome in many scientific endeavors. One of the most common methods is generalized propensity scores (GPS) that extends standard propensity score \citep{rubin1990comment} from a binary treatment to the multiple treatment setting. GPS has been widely used in many causal inference models such as inverse probability of treatment weighted \citep{mccaffrey2013tutorial}, propensity score matching \citep{dehejia2002propensity,rassen2011simultaneously}, subclassification \citep{rosenbaum1984reducing}, and imputations \citep{gutman2015estimation}. For example, \citet{zanutto2005using} extended previous work on evaluating a national anti-drug media campaign into multiple-treatment doses using propensity score subclassification. Genome-wide association studies (GWAS) are typical settings of multiple causal inference, where the multiple treatments are in the form of genetic variations across multiple sites \citep{yu2006unified,ranganath2018multiple}. Multiple causal inference also requires new kind of assumptions, cf.\ the discussions of the assumptions for effect identification and estimation by \citet{lechner2001identification}. 

The aforementioned approaches, however, overlook the presence of hidden confounding bias. In computational genetics, a variety of methods have been proposed to account for hidden confounders in GWAS \citep{yu2006unified,song2015testing}. For instance, \citet{tran2017implicit} described an implicit causal model that leverages neural architectures with an implicit density for GWAS in the presence of hidden confounders. The recently proposed Deconfounder~\citep{wang2019blessings} used the dependencies among multiple treatment as the indirect evidence to find a substitute confounder. It is attractive because of its simplicity and amenability to predictive validation. Our work leverages the advantages of Deconfounder to tackle the challenges in automated meta-analysis, for which we provide unique understandings from the multiple-causal-inference perspective.
\begin{figure}[t]
\centering
    \includegraphics[width=.8\columnwidth]{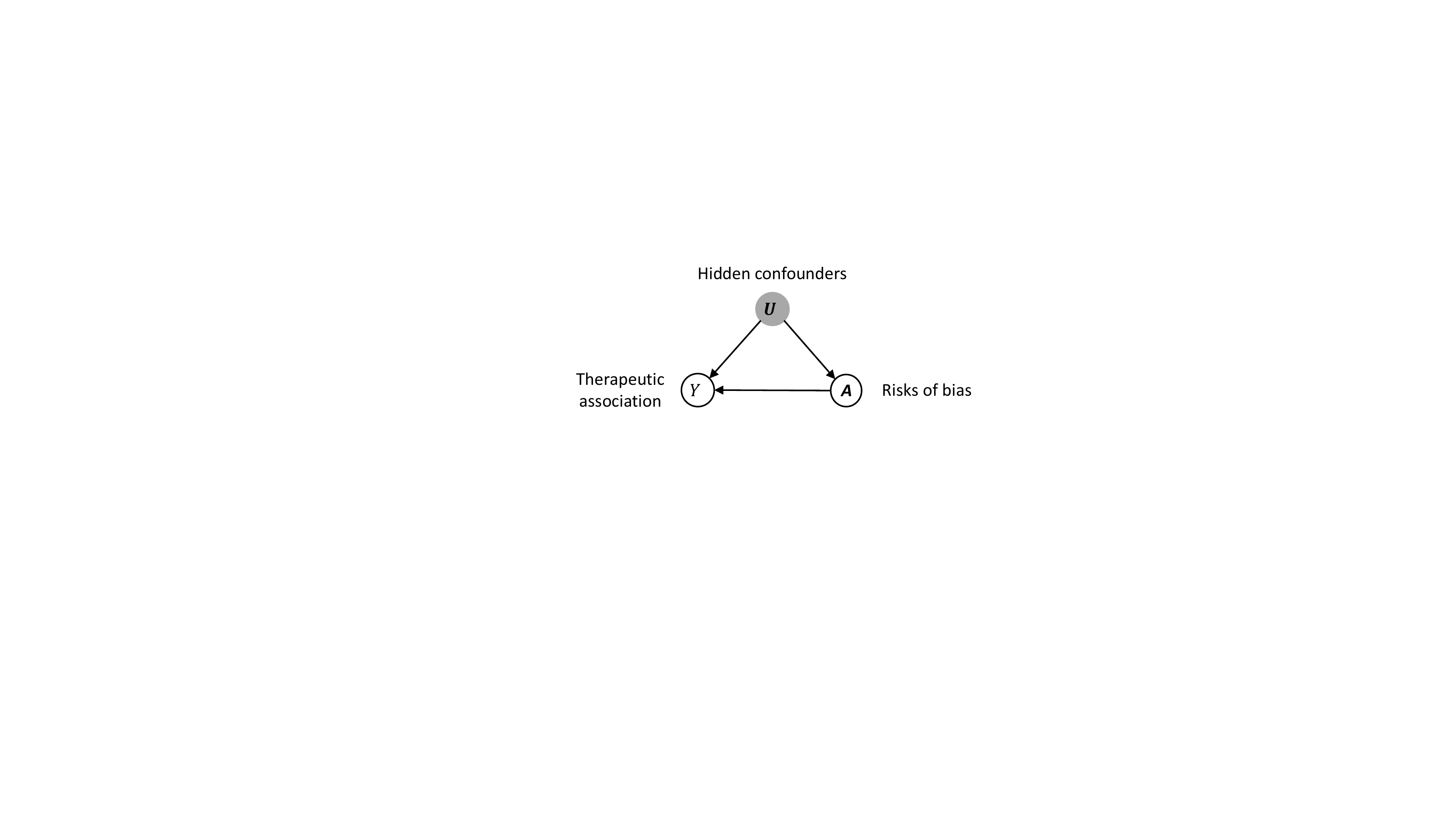}
  \caption{Illustration of multiple causal inference in the presence of hidden confounders for automated meta-analysis.} 
  \label{causal-graph}
\end{figure}

\section{Problem Description}
How can one effectively estimate the summary therapeutic association based on the outputs of these NLP systems? The major obstacle is that neither therapeutic association nor risks of bias belong to the standard input of conventional meta-analysis models. In addition, the potential biases in RCTs, the uncertainty in NLP systems, and other variables that we cannot measure or even we do not realize their existence further induce undesired factors into the procedure of predicting summary therapeutic association.

These unique challenges motivate us to formulate the problem from a causal inference perspective. In particular, we aim to uncover the underlying causal mechanisms among the risks of bias, therapeutic association, and unmeasured factors. The anchor of knowledge we leverage here is that conflicting therapeutic associations we observe across different RCTs are, in part, \textit{caused} by the measured risks of bias. The causal relationship can be further confounded by unmeasured variables, or hidden confounders. For example, analysts' domain expertise can cause them to selectively report the outcome and the summary therapeutic association; uncertainty (e.g., potential errors) in NLP systems can also influence the results of extracted risks of bias and therapeutic association. With \textit{causality-centered} meta-analysis, we answer the question: \textit{What will the therapeutic association be if all risks of bias become zeros?} Or formally, we seek to infer the summary therapeutic association by \textit{intervention}: summary therapeutic association can surface when we intervene on the risks of bias so that their influence on therapeutic association is eliminated. 
\section{Proposed Framework}
\label{sec:framework}
In this section, we introduce the proposed multiple causal inference for automated meta-analysis (MCMA). MCMA employs existing NLP systems for extraction of risks of bias and therapeutic association, which are then used to estimate the summary therapeutic association across several RCTs.

\subsection{Automatic Data Extraction}
The NLP system we use for extracting therapeutic association and risks of bias assessment is the RobotReviewer pretrained on 12,808 RCTs with ground-truth labels \citep{marshall2016robotreviewer}. The RobotReviewer takes a (PDF-formatted) RCT publication as input and labels it with its estimated therapeutic association $Y \in \{0,1,2\}$ (negative, no, and positive association). Another important task it performs is estimating RCT's risks of bias in $D = 6$ domains predefined by the Cochrane risk of bias tool \citep{higgins2011cochrane}: 
\begin{itemize}[noitemsep]
    \item Random sequence generation;
    \item Allocation concealment;
    \item Blinding of participants and personnel;
    \item Blinding of outcome assessment;
    \item Incomplete outcome data;
    \item Selective outcome reporting.
\end{itemize}
For each domain, RobotReviewer determines whether an RCT is at low or high risk of bias. What we need for the downstream summary therapeutic association estimation is the predicted therapeutic association $Y$ and a matrix $\mat{A} \in \{0,1\}^D$, where 0/1 denotes a low/high risk of bias. Note that we recognize the importance of observational studies, however, due to the fact that RobotReviewer is designed for RCTs, our work here focuses on meta-analysis with RCTs. 

\subsection{Multiple Causal Inference for Automated Meta-Analysis}
Essentially, the summary therapeutic association refers to the therapeutic association when no risks of bias are observed in any of the six domains. Underpinning this is the assumption that the risks of bias are the causes of the inconsistent therapeutic associations across multiple RCTs related to the same treatment and outcome. One may directly employs an off-the-shelf multi-class classifier (e.g., multinomial logistic regression) and predict the summary therapeutic association by setting risks of bias to zeros. However, this method has a number of flaws: 
\begin{enumerate}[noitemsep]
    \item \textit{Out-of-Sample Prediction.} In the training data, we do not have access to RCTs where there is no risks of bias. Therefore, directly using trained multi-class classifiers to predict therapeutic association with zero risks of bias can result in unsatisfactory performance. 
    \item \textit{Unmeasured Factors.} In addition to the risks of bias, other unmeasured factors can also lead to conflicting therapeutic associations we observe in different RCTs, e.g., the medical researcher's intention to assign the treatment to certain type of patients in order to amplify the treatment effect. These factors might also influence how we measure risks of bias, e.g., domain expertise causes the bias of selective outcome reporting. 
    \item \textit{Uncertainty in NLP Systems.} NLP systems can make mistakes, especially when they are applied to data collected from specialized domains. 
\end{enumerate}
We propose to address these issues by leveraging the power of multiple causal inference. In contrast to multi-class classifiers aimed at prediction, causal learning models seek to explain and understand the underlying data generating process \citep{pearl2019seven,guo2020survey}. They have superior robustness or adaptability compared to machine learning models \citep{pearl2019seven}. The goal is now to uncover the causal relationship between risks of bias, i.e., the multiple treatments, and therapeutic association, i.e., the outcome, in the presence of hidden confounders $\mat{U}$. Fig.~\ref{causal-graph} characterizes the causal mechanisms among $\mat{A}$, $Y$ and $\mat{U}$.

Given $\mat{A}$ and $Y$ extracted from a corpus of $N$ RCTs $\mathcal{C}$, we infer the summary therapeutic association with $\bm{a}=\textbf{0}$, i.e., $Y(do(\bm{a}=\textbf{0}))$, where $do(\cdot)$ represents the intervention \citep{imbens2015causal,pearl2009causal}. We first infer the mapping function $f(\cdot)$ that characterizes $\mathbb{E}[Y_i(\bm{a})]$ for $i\in \{1,2,...,N\}$ and for each configuration of $\bm{a}\in \mathcal{A}$, where $\mathcal{A}$ denotes the space of all possible risks of bias:
\begin{equation}
     p(Y|do(\bm{a}))\Leftrightarrow f: \{0,1\}^D \xrightarrow{do(\cdot)} \{0,1,2\} \quad\forall \bm{a}\in \mathcal{A}.
\end{equation}
An important notion is that $p(Y|\bm{a})$ indicates the \textit{correlation} whereas $p(Y|do(\bm{a}))$ indicates that the change of $Y$ is the result of the change of $\bm{a}$, i.e., the \textit{causation}.

\subsection{Inference Algorithm}
We reduce our research question to predicting therapeutic association under intervention in the setting of multiple causal inference. Toward this end, we leverage the Deconfounder \citep{wang2019blessings}, a recently proposed multiple causal inference model that combines latent-variable models with predictive model checking to control for hidden confounding. At its core, Deconfounder discovers the substitute confounder $\mat{Z}$ from the dependencies among multiple treatments to approximate the hidden confounder $\mat{U}$. 

Suppose that each RCT is represented by a vector $\bm{a}=\{a_1, a_2, ..., a_D\}$. A potential outcome function $y_i(\bm{a}): \{0,1\}^D \rightarrow \{0,1,2\}$ maps configurations of risks of bias to the outcome therapeutic association for each RCT $i$. The goal of multiple causal inference is to characterize the sampling distribution of the potential outcome, i.e., $\mathbb{E}[Y_i({\bm{a}})]$. We do not have access to the full distribution of $Y_i(\bm{a})$ for any $\bm{a}$ due to the fundamental problem of causal inference \citep{holland1986statistics} that we can only observe one potential outcome (either outcome under treated or control) for each subject. Basic classifiers directly estimate the conditional distribution $\mathbb{E}[Y_i(\bm{a})|\mat{A}=\bm{a}]$. However, in the presence of hidden confounders, $\mathbb{E}[Y_i(\bm{a})]\neq \mathbb{E}[Y_i(\bm{a})|\mat{A}=\bm{a}]$. We first elaborate the required assumptions: 
\begin{enumerate}[noitemsep]
    \item \textit{Stable Unit Treatment Value Assumption (SUTVA)} \citep{rubin1980randomization,rubin1990comment}. There is no interference between individuals and one single version of each cause.
    \item\textit{Positivity}. The substitute confounder $\mat{Z}_i$ satisfies:
    \begin{equation}
        p(A_{ij}\in \mathcal{A}|\mat{Z}_i)>0, \quad p(\mathcal{A})>0,
    \end{equation}
    where $\mathcal{A}$ is the set of $A_{ij}, i=1,2...,N, j=1,2,...,D.$
     \item \textit{No unobserved single-cause confounders}. Formally,
    \begin{equation}
        A_{ij}\independent Y_i(\bm{a})|\mat{X}_i, \quad j=1,...,D,
    \end{equation}
    where $\mat{X}_i$ is some observed background variable.
\end{enumerate}
\noindent The SUTVA assumption implies that the risks of bias of one RCT do not affect risks of bias and therapeutic association of any other RCT. The positivity assumption indicates that given the substitute confounder, the probability of risks of bias being high should be non-zero in each domain. The last assumption, also referred to as \textit{single-ignorability}, implies that there are no such hidden confounders that exclusively influence one single domain of risk of bias. While this assumption is also untestable in practice, it is weaker than assuming there is no hidden confounder. 

Given $\mat{A}$ and $Y$, MCMA consists of three major steps: substitute confounder inference, predictive check of the latent-variable model, and outcome model inference. First, we examine the correlations of all pairs of risks of bias, and remove highly correlated ones to better satisfy the single-ignorability assumption. Next, we define and fit a latent-variable model of the risks of bias: $p(\mat{z},a_1,a_2,...,a_D)$, where $\mat{z}\in \mat{Z}$. Specifically, the model is characterized as 
\begin{equation}
\small
\begin{split}
     \mat{Z}_i\sim p(\cdot|\alpha)\quad i=1,...,N, \\
     A_{ij}|\mat{Z}_i\sim p(\cdot|\mat{z}_i,\theta_j)\quad j=1,...,D,
\end{split}
\end{equation}
where $\alpha$ and $\theta_j$ parameterize the distribution of $\mat{Z}_i$ and the per-cause distribution of $A_{ij}$, respectively. The latent-variable model for the experimentation is the probabilistic PCA (PPCA) \citep{tipping1999probabilistic}. Other factor models are left to explore in future work.
\begin{theorem}
If the distribution of the observed risks of bias $p(\bm{a})$ can be written as the factor model $p(\theta_{1:D},\bm{z},\bm{a})$, we obtain the following ignorable treatment assignment:
\begin{equation}
    Y_i(\bm{a})\independent (\mat{A}_{i1},...,\mat{A}_{iD})|\bm{Z}_i \quad \forall i\in\{1,2...,N\}.
\end{equation}
\end{theorem}
\noindent This theorem is proved by the fact that the substitute confounder $\bm{Z}$ is inferred without knowledge of the potential outcome $Y$ and the fact that the risks of bias $(\mat{A}_{i1},...,\mat{A}_{iD})$ are jointly independent given $\bm{Z}$. The results indicate that $\bm{Z}$ captures all the dependencies between the therapeutic association and the risks of bias. The identifiability holds if all the required assumptions are satisfied. When the effects are not identifiable, the estimates will present large variance and uncertainty in the potential outcomes can be used to quantify the reliability of the inference algorithm.

In the second step of predictive check, we randomly hold out a subset of the risks of bias for each RCT $i$, denoted as $\bm{a}_{i,held}$ and the rest $\bm{a}_{i,obs}$. We then fit the latent model to $\{\bm{a}_{i,obs}\}_{i=1}^N$ and perform a predictive check on the held-out dataset. A predictive check compares the given risk of bias with the risk of bias drawn from the model's predictive distribution \citep{wang2019blessings}. If the predictive check score $p\in(0,1)$ is larger than 0.5, we conclude that the latent model generates values of the held-out causes that give similar log-likelihoods to their real values. 

In the final step of inferring outcome model, we use the fitted factor model $L$ to infer the substitute confounder for each RCT $p(\mat{z}_i|\bm{a}_i)$, i.e., $\mat{z}_i=\mathbb{E}_L[\mat{Z}_i|\mat{A}_i=\bm{a}_i]$. We then augment the risks of bias $\bm{a}_{i}$ with $\mat{z}_i$. The outcome model can be estimated as follows:
\begin{equation}
    \mathbb{E}[Y_i(\mat{A}_i)|\mat{Z}_i=\mat{z}_i,\mat{A}_i=\bm{a}_i]=f(\bm{a},\mat{z})
\end{equation} 
with augmented data $\{\bm{a}_i,\mat{z}_i,y_i(\bm{a}_i)\}$ via a multi-class classification model $f(\bm{a},\mat{z})$, e.g., multinomial logistic regression. At last, we infer the summary therapeutic association by setting $\bm{a}=\textbf{0}$ in the fitted outcome model.

We recognize that there have been discussions (e.g., \citep{ogburn2019comment}) about the identification issues with Deconfounder. However, given the fact that we work on data automatically extracted from scientific publications by NLP systems, which is different from conventional data evaluated in causal inference (i.e., we assign treatment and observe outcomes), we consider a causal graph in its simplest form with which the Deconfounder is legitimate to use. While Deconfounder is not appropriate in all situations \citep{athey2019comment}, we show the validity of using Deconfounder in our problem setting in Appendix A. To this end, we expect Deconfounder to eliminate or at least reduce the hidden confounding in automated meta-analysis. 
\begin{table*}[t]
\small
\setlength\tabcolsep{3pt}
\begin{center}
\caption{Performance comparisons w.r.t.\ predicting observed therapeutic associations.}
\begin{tabular}{ abababababab } \hline
\rowcolor{white}
\multicolumn{10}{c}{\textbf{AUC scores}}\\\hline
\rowcolor{white}
\multicolumn{2}{c}{MNLogit}&\multicolumn{2}{c}{$k$-NN}&\multicolumn{2}{c}{MLP}&\multicolumn{2}{c}{Gaussian NB}&\multicolumn{2}{c}{XGBoost}\\\cmidrule(lr){1-2} \cmidrule(lr){3-4}\cmidrule(lr){5-6} \cmidrule(lr){7-8}\cmidrule(lr){9-10}
Basic&MCMA&Basic&MCMA&Basic&MCMA&Basic&MCMA&Basic&MCMA\\\hline
.507$\pm$.014&\textbf{.523$\pm$.022}&.519$\pm$.018&\textbf{.529$\pm$.024}&.535$\pm$.020&\textbf{.539$\pm$.019}&.573$\pm$.026&\textbf{.581$\pm$.025}&\textbf{.539$\pm$.022}&.536$\pm$.033\\[0.7mm]
\multicolumn{10}{c}{\textbf{F1 scores}}\\\hline
Basic&MCMA&Basic&MCMA&Basic&MCMA&Basic&MCMA&Basic&MCMA\\\hline
.411$\pm$.050&\textbf{.434$\pm$.038}&.421$\pm$.028&\textbf{.433$\pm$.039}&.436$\pm$.032&\textbf{.437$\pm$.035}&\textbf{.424$\pm$.037}&.421$\pm$.037&.423$\pm$.035&\textbf{0.443$\pm$0.043}
\end{tabular}
\label{outcome}
\end{center}
\end{table*}
\section{Empirical Evaluation}
Empirical evaluation for automated meta-analysis on real-world data is extremely challenging because: (1) we can only measure one potential outcome for the same subject in the same study; (2) the number of RCTs studying the same treatment/risk factor and disease is limited due to the ethical and financial considerations when conducting RCTs; and (3) we need to follow the standard meta-analysis process, which is extremely time-consuming, to obtain the ground truth for the summary therapeutic association. Consequently, to examine the different aspects of MCMA, here, we follow standard causal inference evaluation method and focus on synthetic data for which we know the ground truth. In addition, we will also present a case study using semi-synthetic data generated based on the statistics collected from a real-world meta-analysis. 

The evaluation centers on the following four perspectives: \\
\noindent\textit{P1}. With embedded causal knowledge, will MCMA hurt the performance of standard multi-class classifiers w.r.t. predicting therapeutic association in \textit{individual} RCTs?\\ 
\noindent\textit{P2}. What is the effect of varying the level of introduced confounding noise on model performance w.r.t. predicting the therapeutic association in \textit{individual} RCTs?\\
\noindent\textit{P3}. How does MCMA fare against standard multi-class classifiers w.r.t. the precision of predicting the \textit{summary} therapeutic association?\\
\noindent\textit{P4}. Will the answer to \textit{P1} remain the same using semi-synthetic data generated based on real-world RCTs?

\subsection{Experimental Setup}
We are not aware of any similar work in the literature to automate meta-analysis with risks of bias and hidden confounders. Inaccessible to the background variables $\mat{X}$, conventional multiple causal inference models (e.g., \citep{hill2011bayesian}) under Unconfoundedness assumption\footnote{All the confounders can be captured by $\mat{X}$.} \citep{rubin1980randomization} revert to basic classification models. Thus, given the studied problem inherently relates to multi-class classification, we compare MCMA with widely-used classifiers, including $k$-Nearest Neighbors ($k$-NN), Multinomial Logistic Regression (MNLogit), Gaussian Na\"ive Bayes (Gaussian NB), Multilayer Perceptron (MLP), and XGBoost \citep{chen2016xgboost}. Note that for each baseline, there is a corresponding MCMA that uses the same classification model but integrated with the introduced causal knowledge. We consider three evaluation metrics: AUC and F1 scores for predicting therapeutic association in each RCT, and absolute error of predicting the probabilities of each class being the summary therapeutic association. For the Deconfounder implementation, we used Tensorflow \citep{abadi2016tensorflow} and Statsmodels \citep{seabold2010statsmodels}. The dimension of the substitute confounder $\mat{Z}$ is set low (e.g., 1 in our implementation) to keep the estimation variance small \citep{wang2019blessings}. To enforce positivity, it is suggested that the dimension of $\mat{Z}$ is set to be smaller than the number of treatments in practice \citep{wang2019blessings}. The latent-variable model PPCA is optimized by Adamax \citep{kingma2014adam} with a learning rate of 0.01. 
\subsection{Synthetic Experiments}
We begin by simulating the causal mechanism described in Fig.\ \ref{causal-graph}. Specifically, we define $\mat{A}$, $Y$ and $U$ as:
\begin{equation}
\small
    \begin{split}
        u_i\sim \text{Unif}(0,1)\\
         \bm{a}_i|u_i \sim \text{Bern}(0.75u_i+0.25(1-u_i));\\
         y_i \sim \text{Multinomial}(p_1,p_2,p_3),\\
         \text{where } p_1=\frac{\mat{w}_1^\intercal\bm{a}_iu_i+4u_i}{l},p_2=\frac{\mat{w}_2^\intercal\bm{a}_i}{l},p_3=\frac{\mat{w}_3^\intercal\bm{a}_i+w_uu_i}{l};\\
         l=\mat{w}_1^\intercal\bm{a}_iu_i+4u_i+\mat{w}_2^\intercal\bm{a}_i+\mat{w}_3^\intercal\bm{a}_i+w_uu_i,\\
         \mat{w}_1\sim \text{Possion}(3),\mat{w}_2\sim \text{Possion}(2),\mat{w}_3\sim \text{Possion}(1).
    \end{split}
    \label{simulation}
\end{equation}
Here, $w_u\in\mathbb{R}$ indicates the level of the induced confounding noise through the hidden confounder $u_i$, which is modeled as a one-dimensional continuous variable following the uniform distribution. We set $w_u=2$ in \textit{P1} and \textit{P3}, and vary $w_u$ in \textit{P2} to examine its impact on model performance. The number of domains for risks of bias is set to $D=10$. We also ensure that at least one domain of risks of bias is high. This data generating process explicitly introduces confounding bias between $\mat{A}$ and $Y$ as they both hinge on the assignment of ``unobserved'' $u_i$. Additionally, we parameterize the distribution of $Y$ in a way that $Y=0$ is more likely to be the summary therapeutic association. By setting $\bm{a}$ to \textbf{0}, we obtain the ground truth for the probabilities of each class being the summary therapeutic association, i.e., $(p^s_1,p^s_2,p^s_3)$. All the experimental results are averaged over 10 replications.

\begin{figure}[t]
\centering
\begin{subfigure}{.5\columnwidth}
\centering
\captionsetup{justification=centering}
  \includegraphics[width=.9\linewidth]{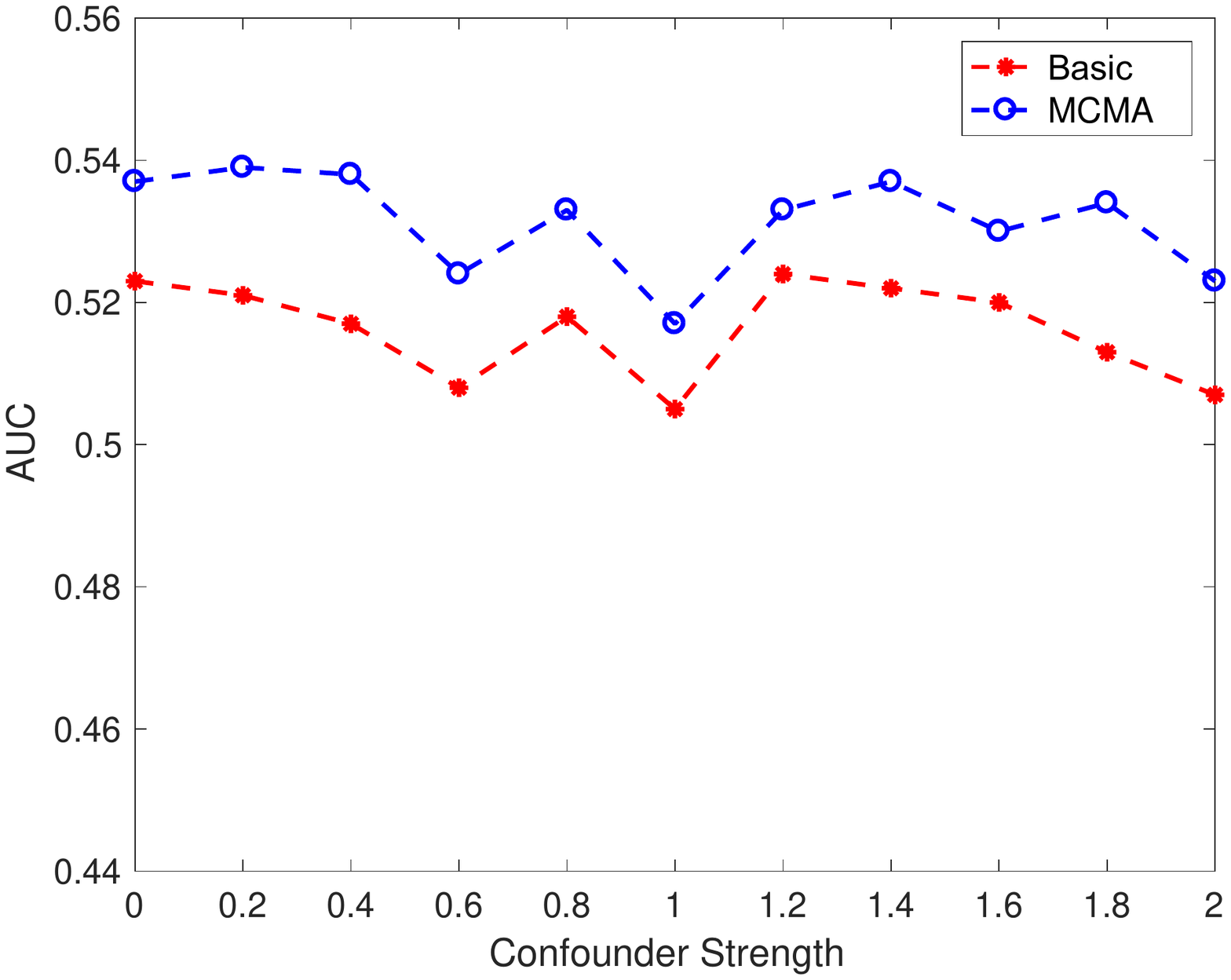}
  \caption{AUC score.}
\end{subfigure}%
\begin{subfigure}{.5\columnwidth}
\centering
\captionsetup{justification=centering}
  \includegraphics[width=.9\linewidth]{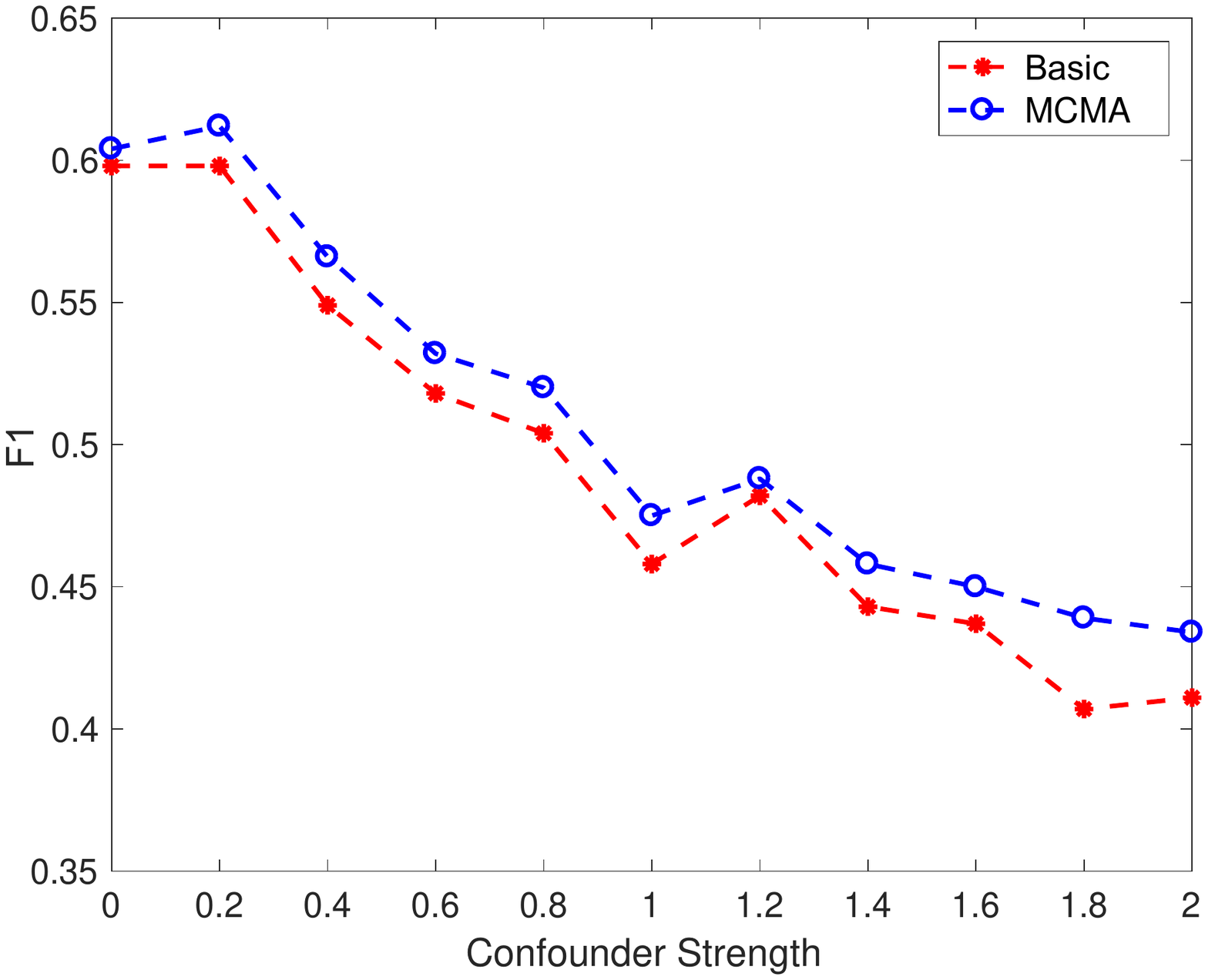}
    \caption{F1 score.}
\end{subfigure}
\caption{Performance comparisons using synthetic data and MNLogit with varied levels of confounding noise.}
\label{confounder}
\end{figure}
\begin{table*}[t]
\small
\setlength\tabcolsep{3pt}
\begin{center}
\caption{Performance comparisons w.r.t. predicting therapeutic association between PDE5 inhibitor and the cardiac morphology in different RCTs.}
\begin{tabular}{ abababababab } \hline
\rowcolor{white}
\multicolumn{10}{c}{\textbf{AUC scores}}\\\hline
\rowcolor{white}
\multicolumn{2}{c}{MNLogit}&\multicolumn{2}{c}{$k$-NN}&\multicolumn{2}{c}{MLP}&\multicolumn{2}{c}{Gaussian NB}&\multicolumn{2}{c}{XGBoost}\\\cmidrule(lr){1-2} \cmidrule(lr){3-4}\cmidrule(lr){5-6} \cmidrule(lr){7-8}\cmidrule(lr){9-10}
Basic&MCMA&Basic&MCMA&Basic&MCMA&Basic&MCMA&Basic&MCMA\\\hline
.500$\pm$.032&\textbf{.504$\pm$.031}&.490$\pm$.071&\textbf{.491$\pm$.050}&.513$\pm$.070&\textbf{.543$\pm$.078}&.512$\pm$.052&\textbf{.521$\pm$.082}&\textbf{.532$\pm$.046}&.520$\pm$.065\\[0.7mm]
\multicolumn{10}{c}{\textbf{F1 scores}}\\\hline
Basic&MCMA&Basic&MCMA&Basic&MCMA&Basic&MCMA&Basic&MCMA\\\hline
.389$\pm$.071&\textbf{.402$\pm$.100}&.365$\pm$.099&\textbf{.384$\pm$.062}&.420$\pm$.078&\textbf{.458$\pm$.089}&\textbf{.410$\pm$.070}&.409$\pm$.085&.427$\pm$.081&\textbf{.434$\pm$.077}
\end{tabular}
\label{PDE5}
\end{center}
\end{table*}
\begin{figure*}[t]
\centering
\begin{subfigure}{.2\textwidth}
\centering
\captionsetup{justification=centering}
  \includegraphics[width=\linewidth]{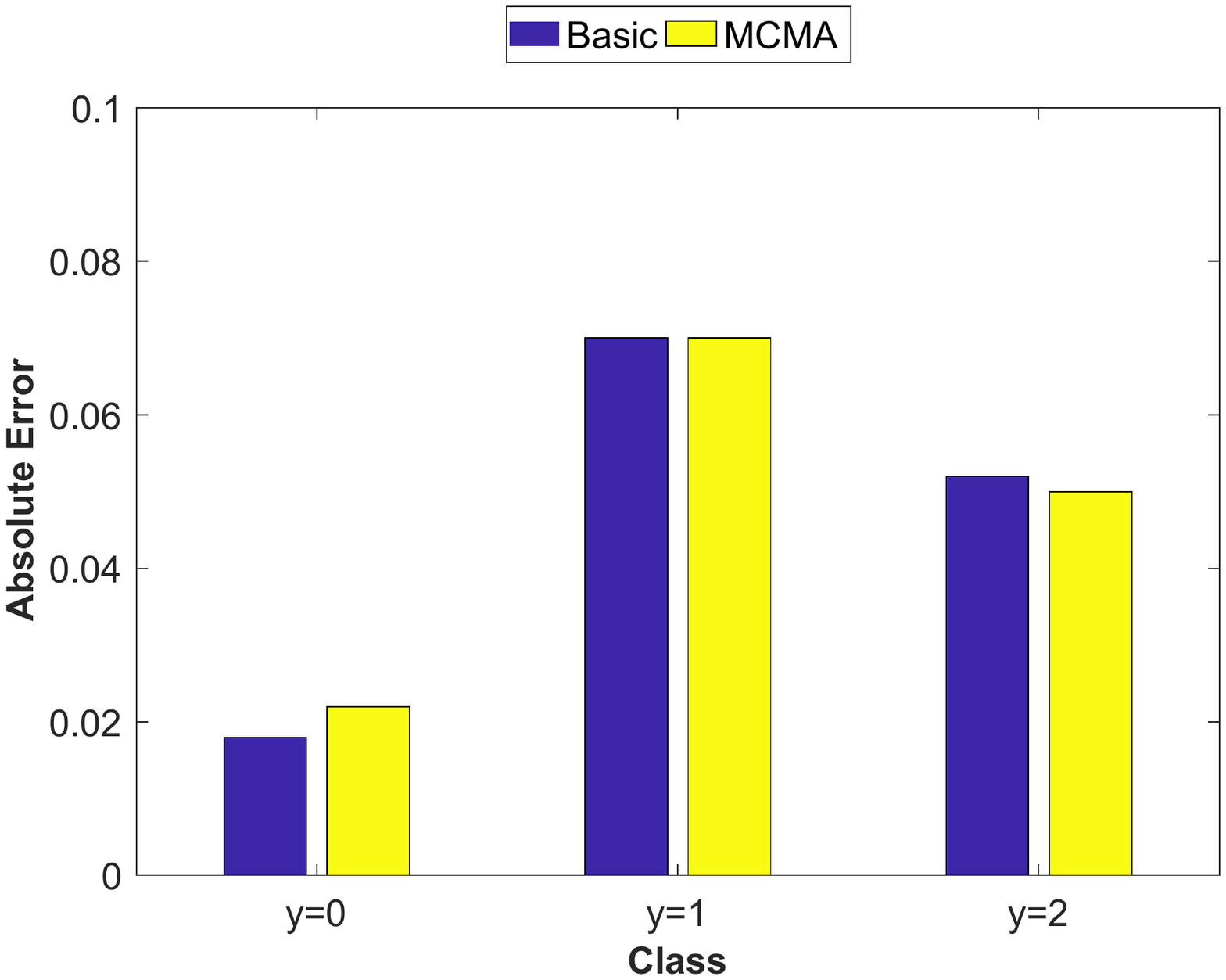}
  \caption{MNLogit.}
\end{subfigure}%
\begin{subfigure}{.2\textwidth}
\centering
\captionsetup{justification=centering}
  \includegraphics[width=\linewidth]{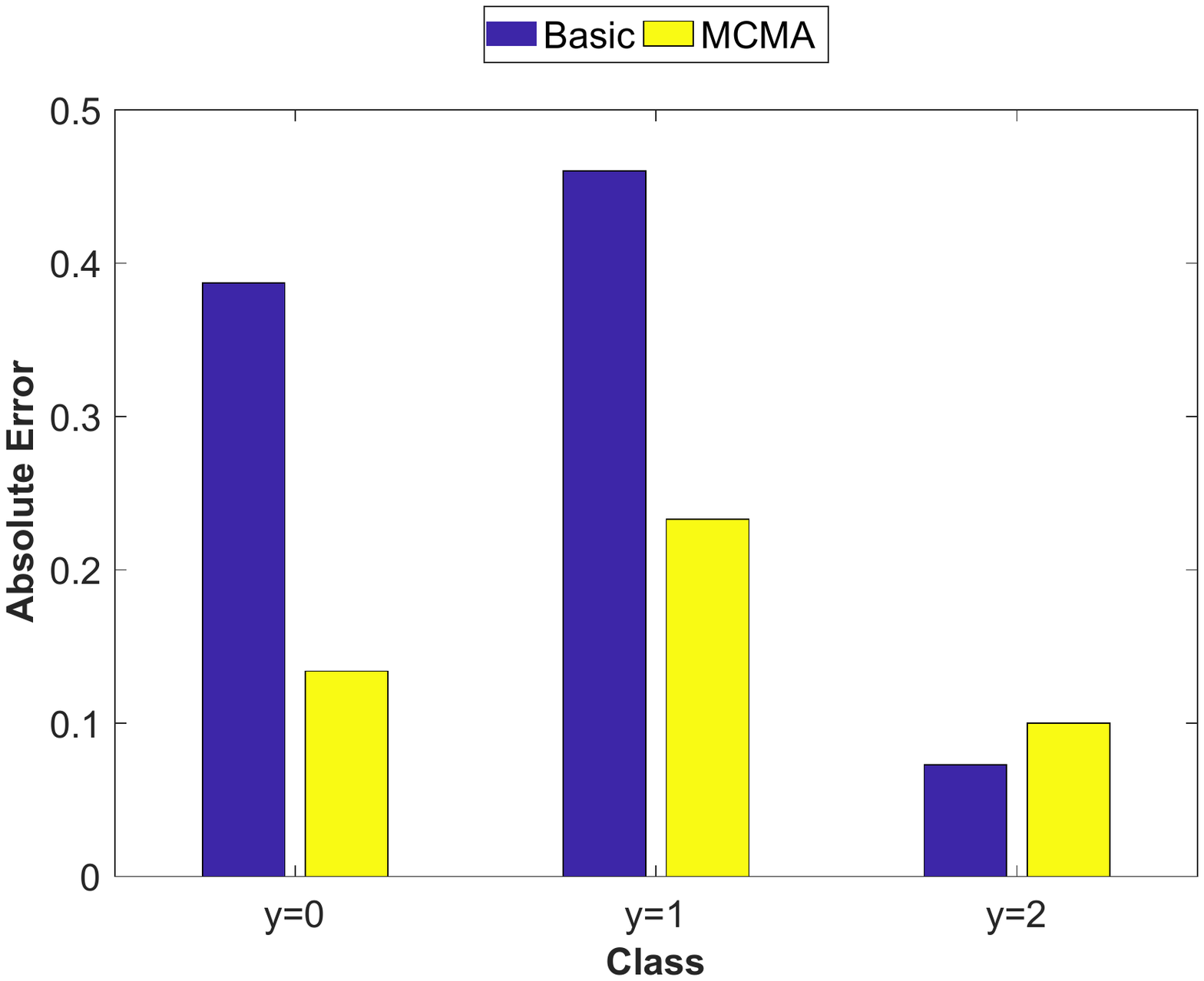}
    \caption{$k$-NN.}
\end{subfigure}%
\begin{subfigure}{.2\textwidth}
\centering
\captionsetup{justification=centering}
  \includegraphics[width=\linewidth]{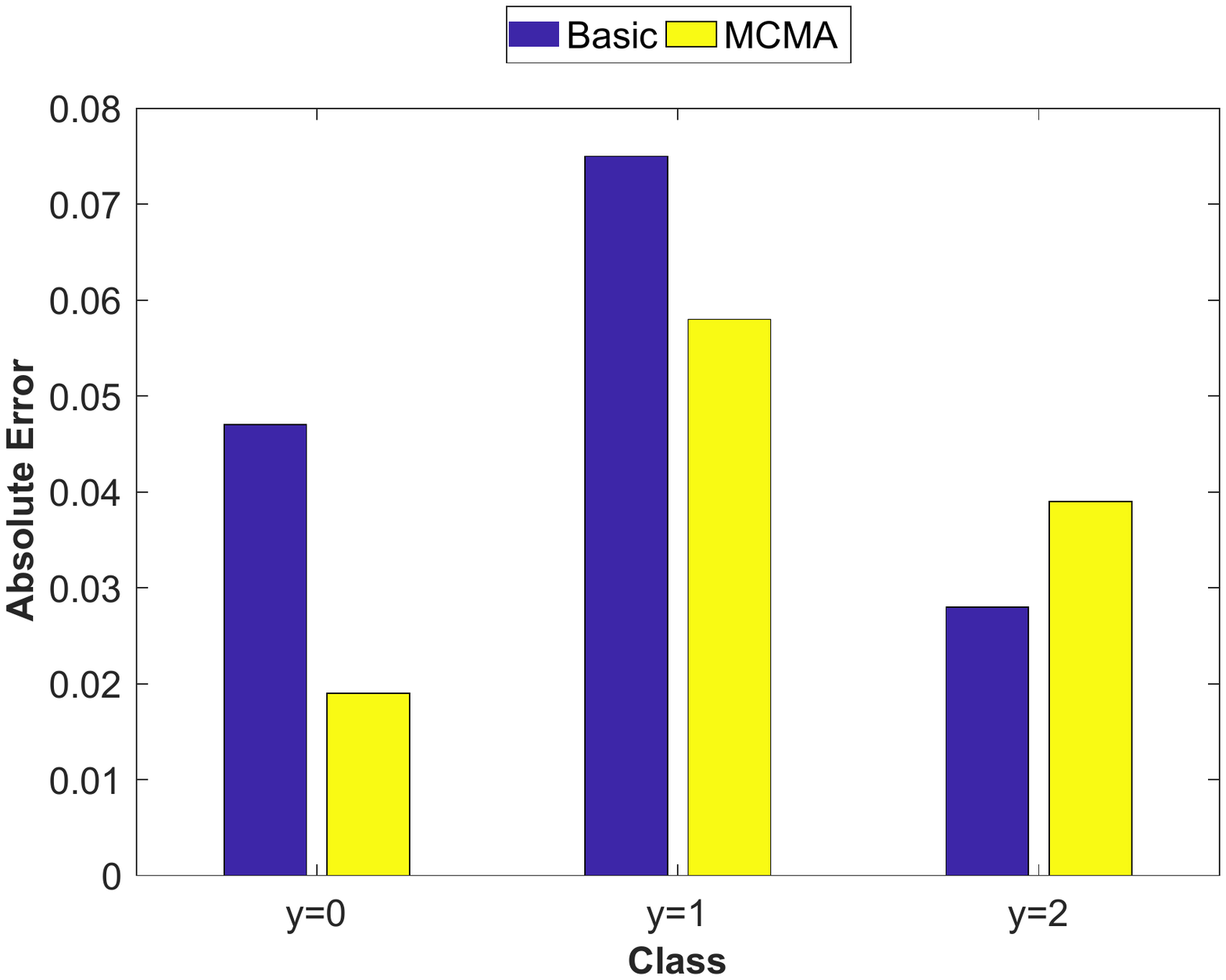}
    \caption{MLP.}
\end{subfigure}%
\begin{subfigure}{.2\textwidth}
\centering
\captionsetup{justification=centering}
  \includegraphics[width=\linewidth]{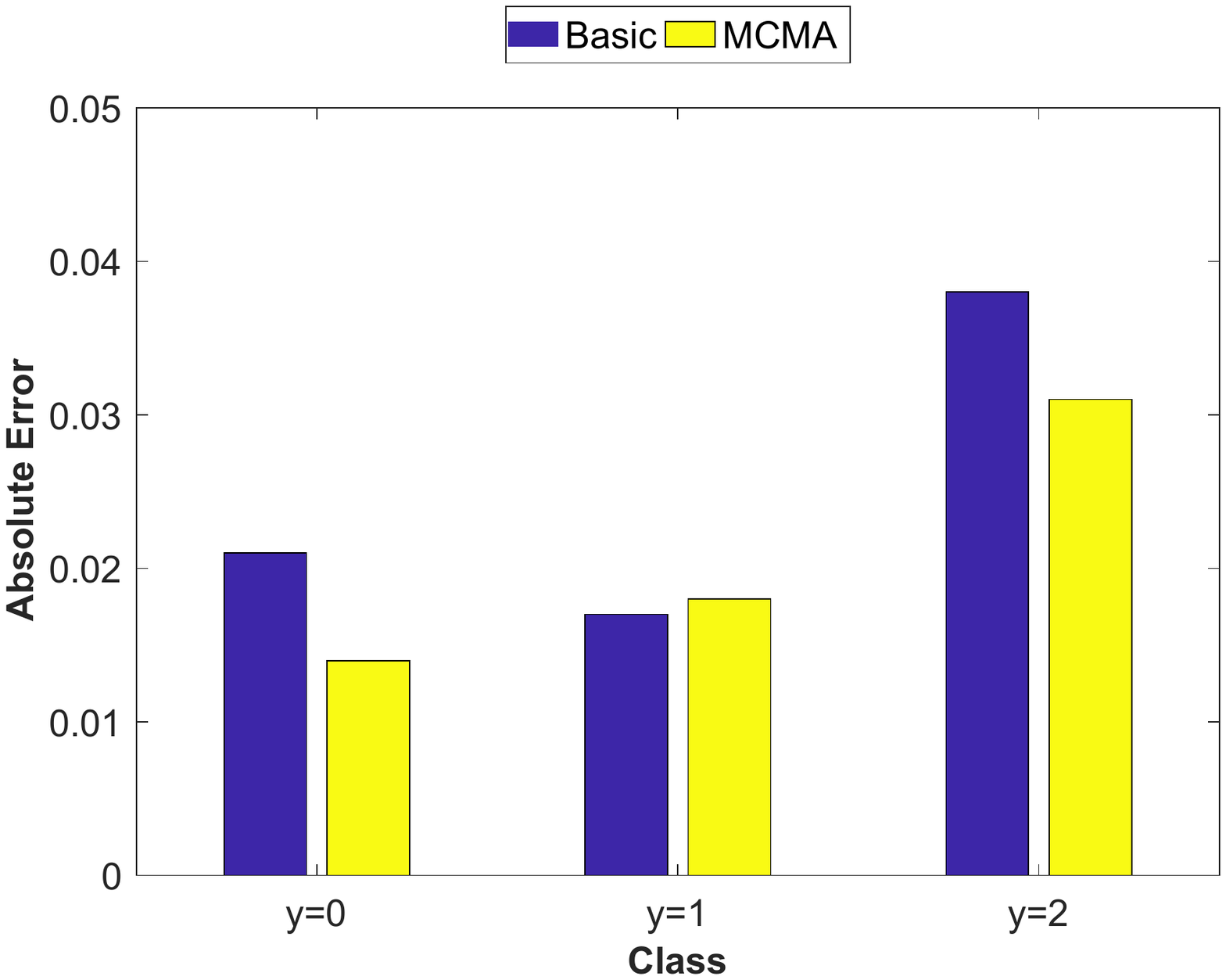}
    \caption{Gaussian NB.}
\end{subfigure}%
\begin{subfigure}{.2\textwidth}
\centering
\captionsetup{justification=centering}
  \includegraphics[width=\linewidth]{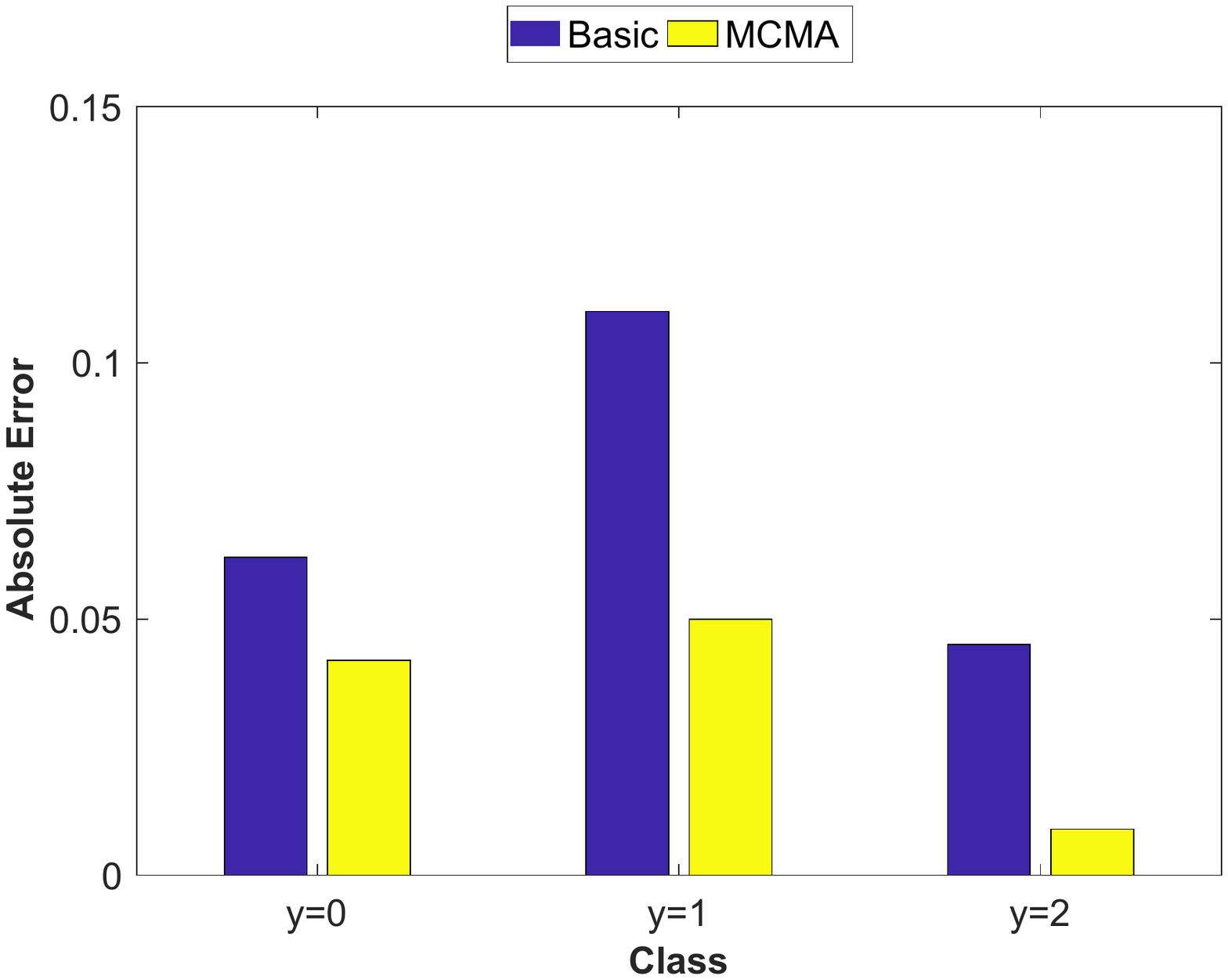}
    \caption{XGBoost.}
\end{subfigure}
\caption{Absolute Errors of the estimated probabilities of being the summary therapeutic association with $N=100$.}
\label{100}
\end{figure*}

\noindent\textbf{Predicting Therapeutic Association in \textit{Individual} RCT (P1).} We first show that, with embedded causal knowledge, MAMC does not hurt the performance on predicting individual therapeutic associations potentially influenced by risks of bias and hidden confounders. We set the sample size $N=1000$ in this experiment and report mean AUC and F1 scores along with standard deviations in Table \ref{outcome}. We denote the baselines as Basic and ours as MCMA. We observe that the causality-guided multi-class classifiers, i.e., MCMA, can achieve competitive performance w.r.t. both AUC and F1 scores. For example, MCMA improves AUC and F1 scores over standard MNLogit by 31.6\% and 56.0\%, respectively. Standard deviations are small overall. 

\noindent\textbf{Varying the Level of Induced Confounding Noise (P2).}
We vary $w_u$ from 0 to 2 with an increment of 0.2 to further examine the effect of the induced confounding noise on models' performance in predicting the therapeutic association. Intuitively, as $w_u$ becomes larger, the increased confounding noise will exacerbate the performance of all models. The generated risks of bias and therapeutic association are then fed into the Basic MNLogit and the corresponding MCMA. Here, we use MNLogit as a working example; similar results are found using other classifiers. Fig.\ \ref{confounder} shows that as expected, the prediction performance of both standard MNLogit and MCMA tend to degrade as we introduce more confounding noise in the data generation process. The trend is especially apparent for F1 score. This is mainly because the synthetic data is simulated to be imbalanced to ensure there are more samples with $Y=0$. As such, the probability of $Y=0$ being the summary therapeutic association will be the largest when $\bm{a}=\textbf{0}$. Compared to AUC, F1 is more sensitive to the effect of the unevenness among classes. Another important finding is that MCMA consistently outperforms Basic MNLogit when varying the confounder strength $w_u$. 

\begin{figure*}[t]
\centering
\begin{subfigure}{.2\textwidth}
\centering
\captionsetup{justification=centering}
  \includegraphics[width=\linewidth]{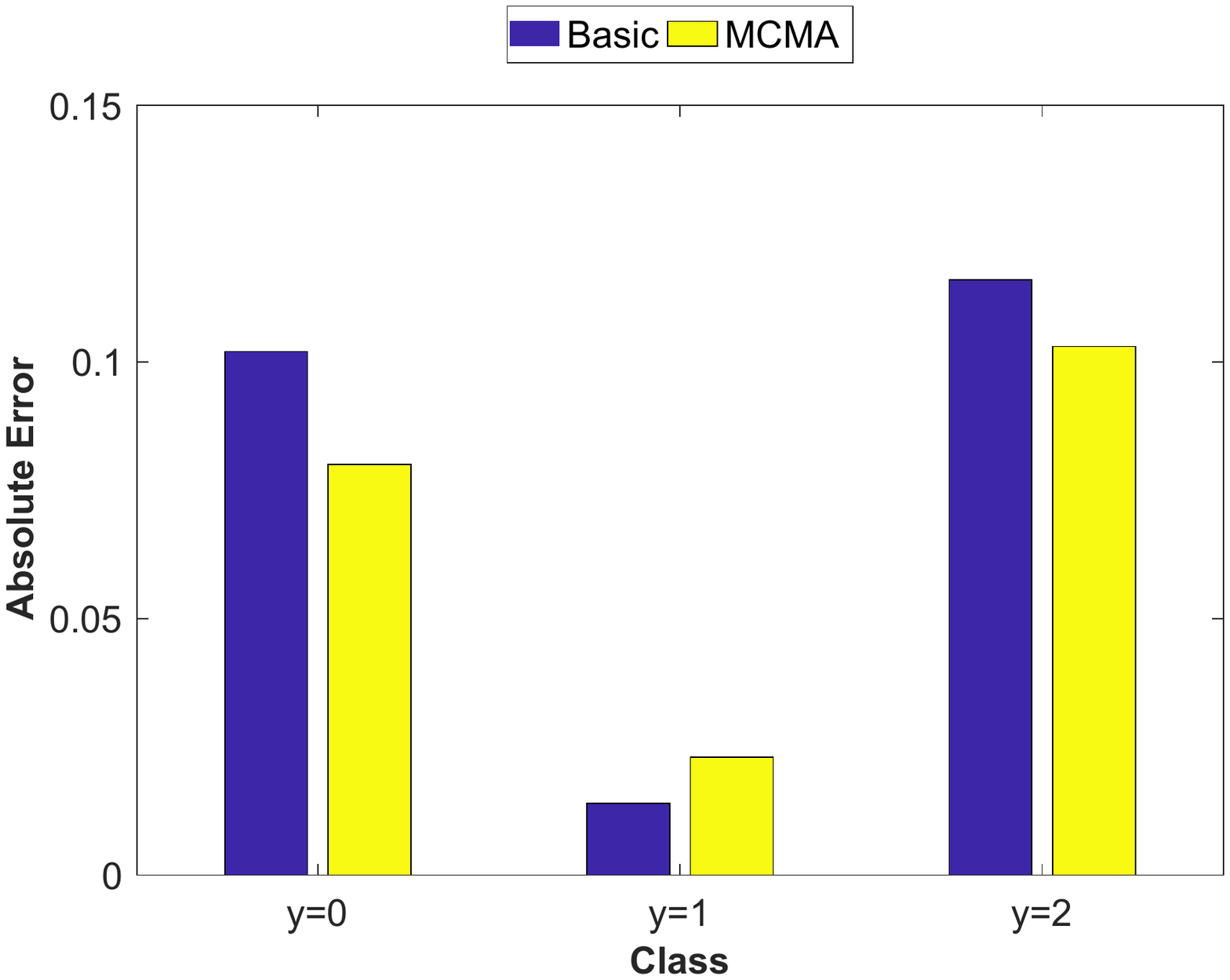}
  \caption{MNLogit.}
\end{subfigure}%
\begin{subfigure}{.2\textwidth}
\centering
\captionsetup{justification=centering}
  \includegraphics[width=\linewidth]{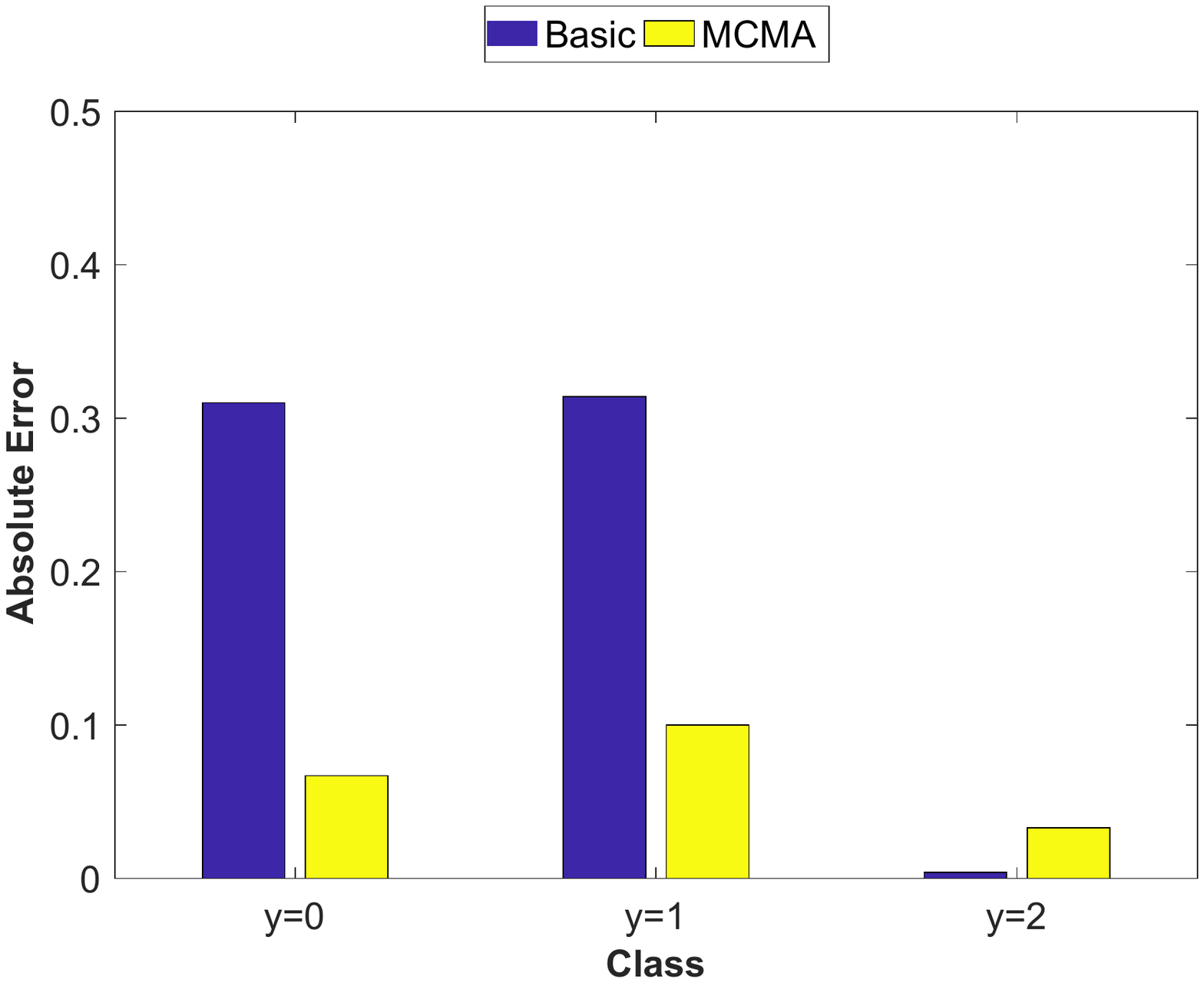}
    \caption{$k$-NN.}
\end{subfigure}%
\begin{subfigure}{.2\textwidth}
\centering
\captionsetup{justification=centering}
  \includegraphics[width=\linewidth]{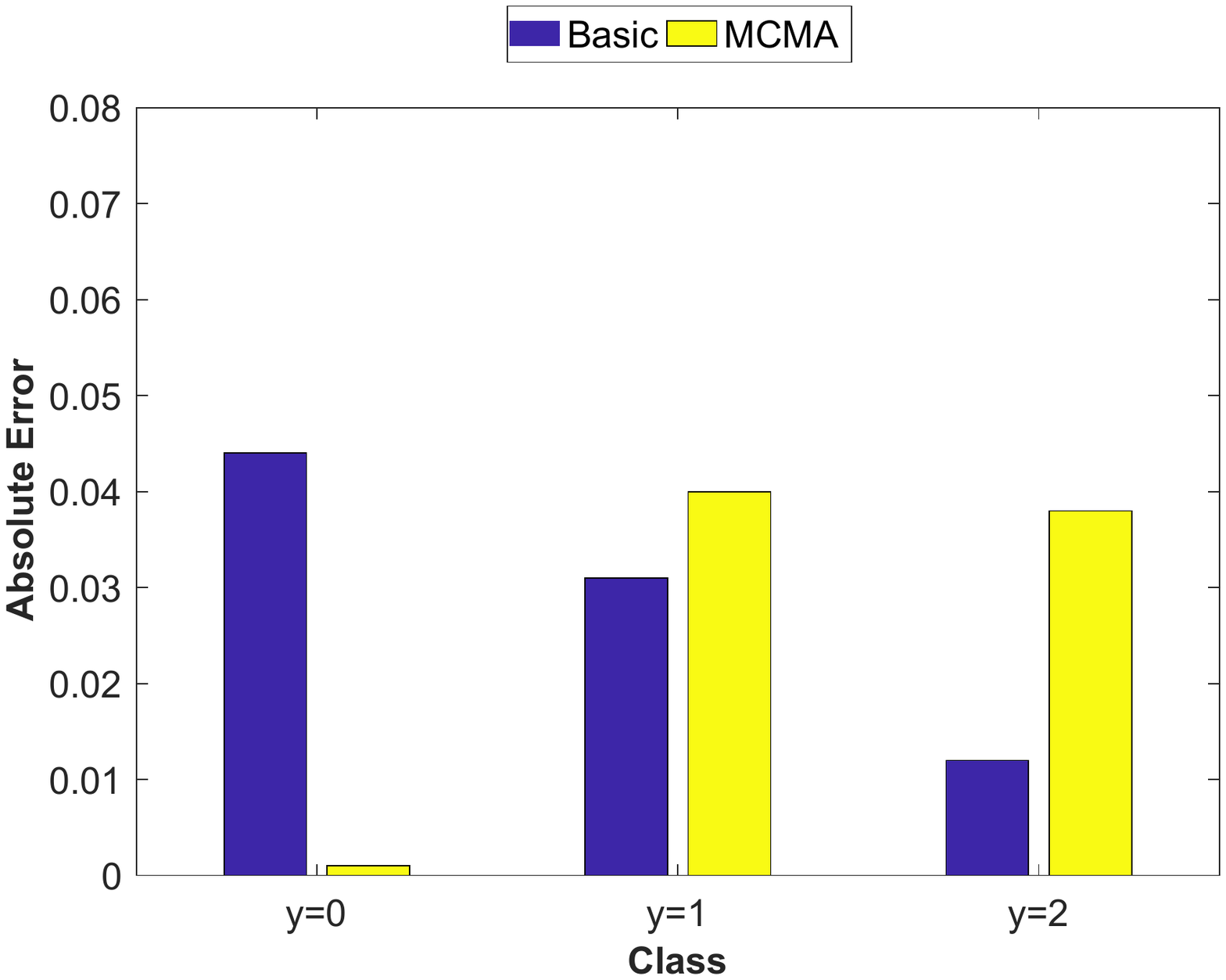}
    \caption{MLP.}
\end{subfigure}%
\begin{subfigure}{.2\textwidth}
\centering
\captionsetup{justification=centering}
  \includegraphics[width=\linewidth]{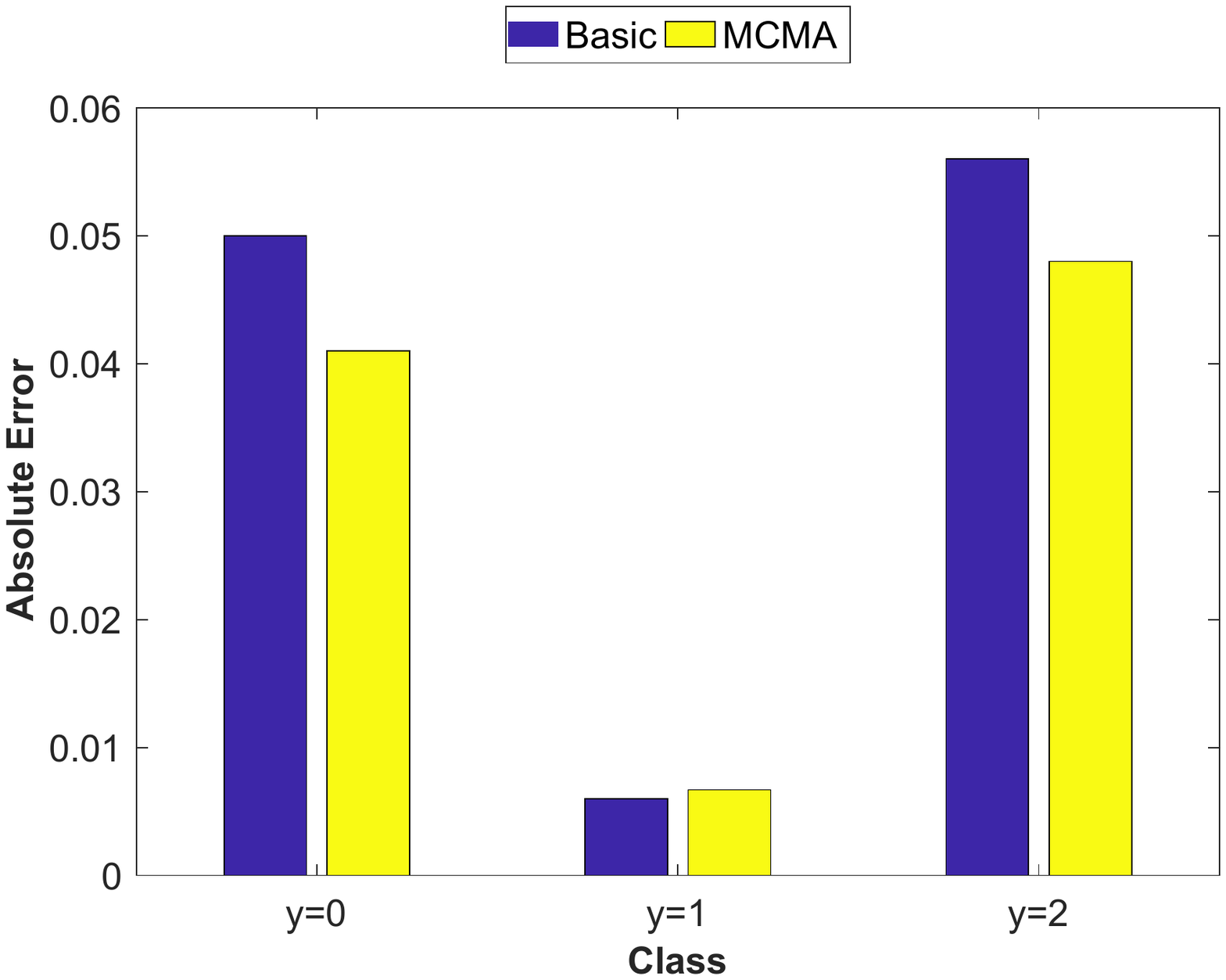}
    \caption{Gaussian NB.}
\end{subfigure}%
\begin{subfigure}{.2\textwidth}
\centering
\captionsetup{justification=centering}
  \includegraphics[width=\linewidth]{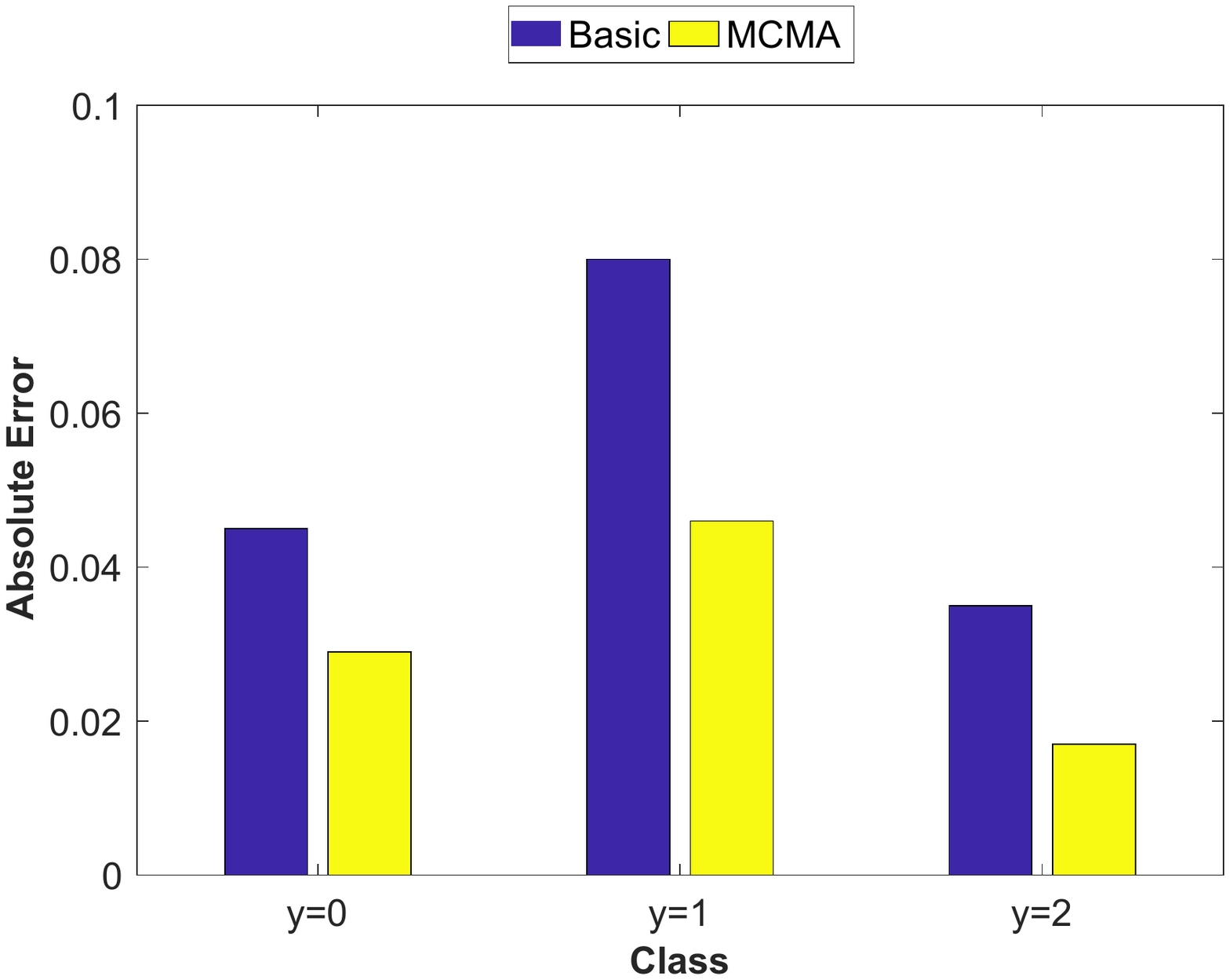}
    \caption{XGBoost.}
\end{subfigure}
\caption{Absolute Errors of the estimated probabilities of being the summary therapeutic association with $N=500$.}
\label{500}
\end{figure*}

\begin{figure*}[t]
\centering
\begin{subfigure}{.2\textwidth}
\centering
\captionsetup{justification=centering}
  \includegraphics[width=\linewidth]{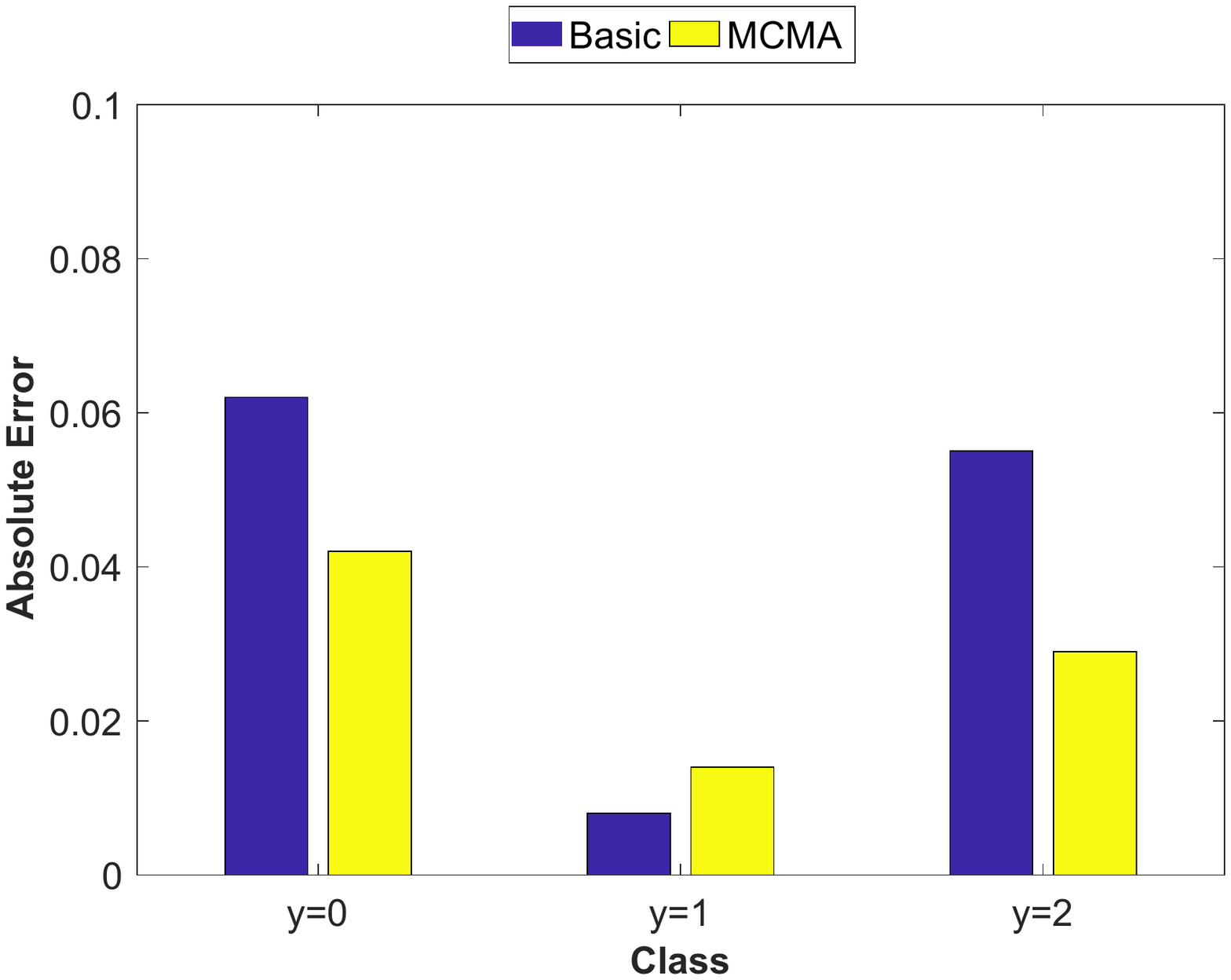}
  \caption{MNLogit.}
\end{subfigure}%
\begin{subfigure}{.2\textwidth}
\centering
\captionsetup{justification=centering}
  \includegraphics[width=\linewidth]{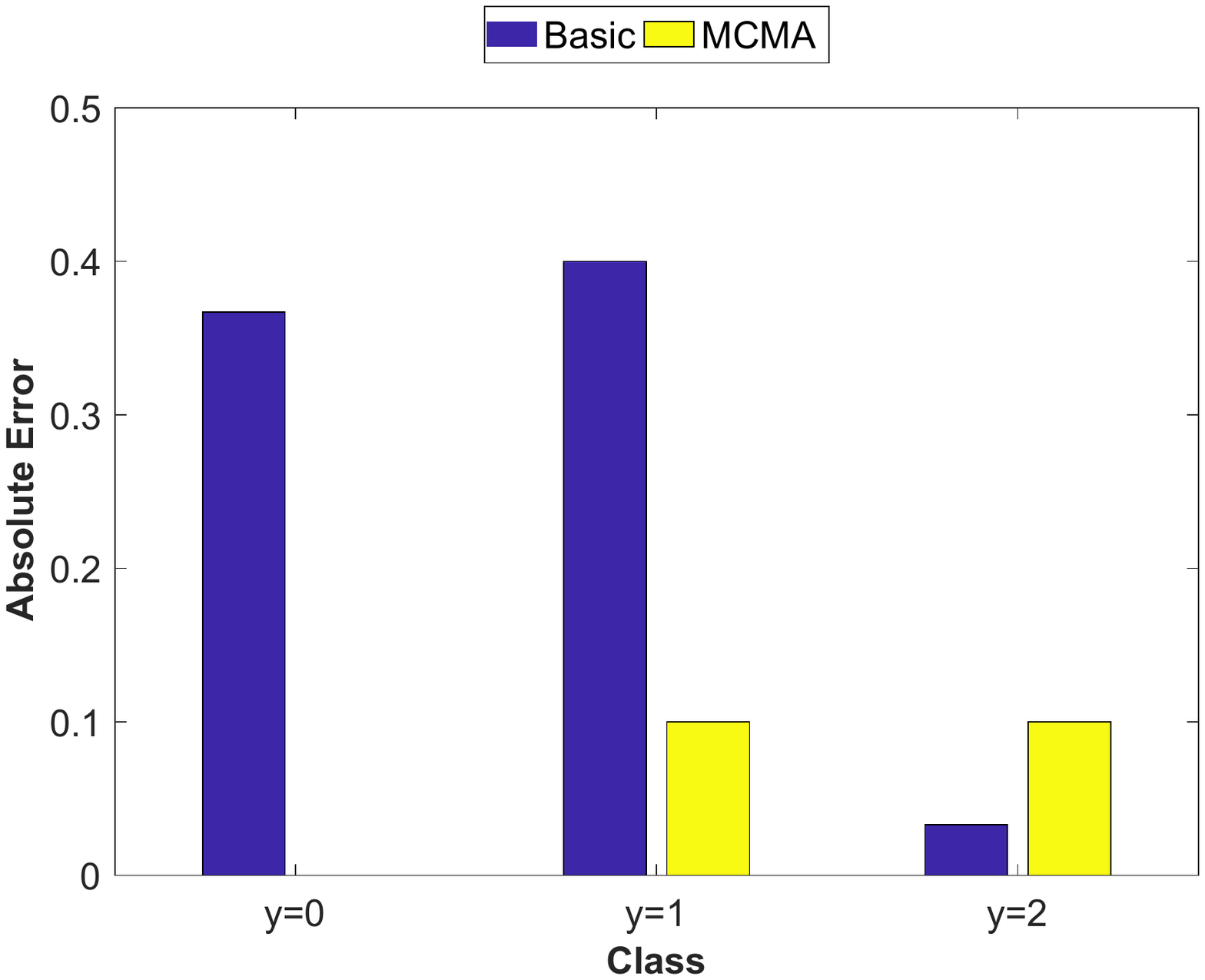}
    \caption{$k$-NN.}
\end{subfigure}%
\begin{subfigure}{.2\textwidth}
\centering
\captionsetup{justification=centering}
  \includegraphics[width=\linewidth]{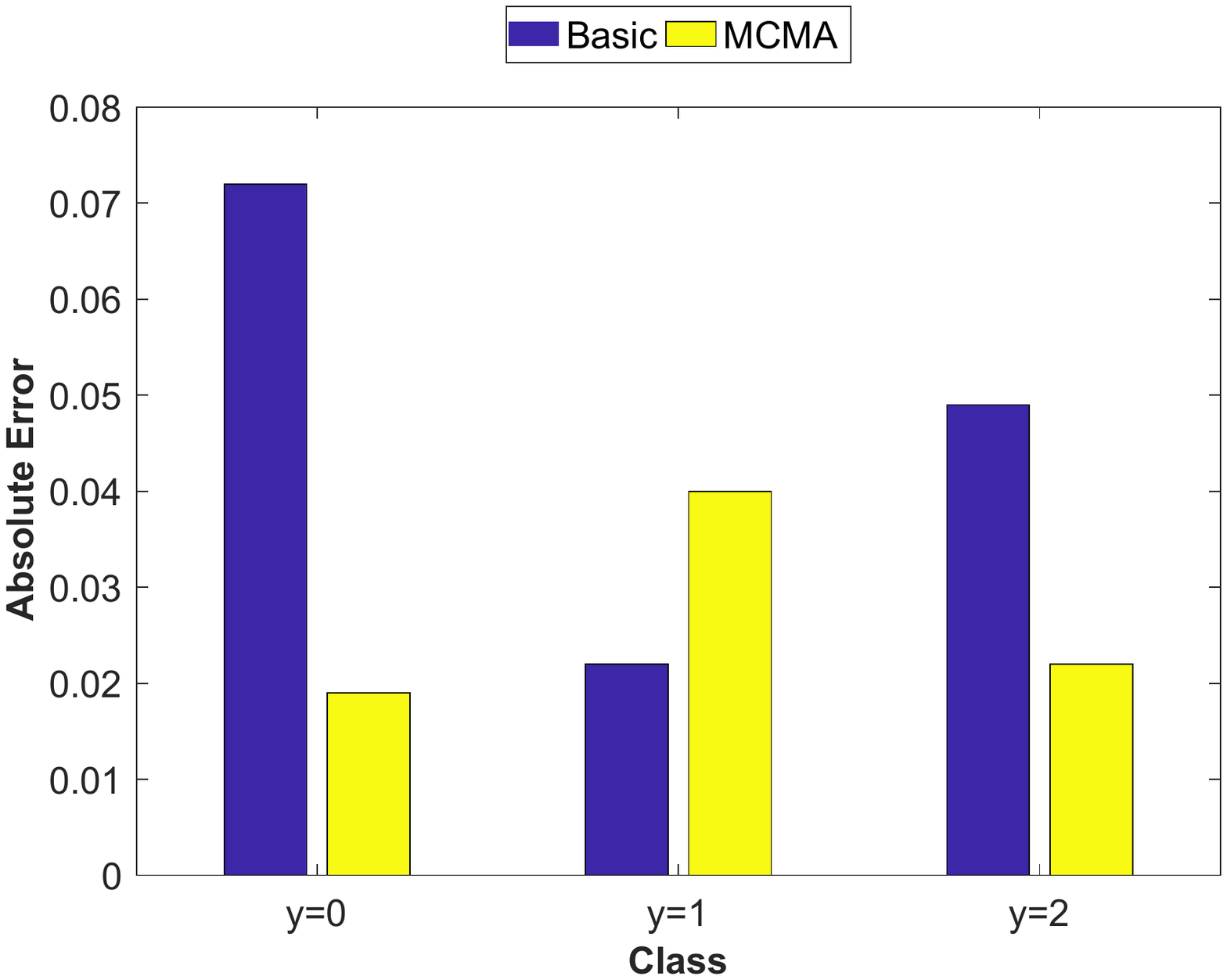}
    \caption{MLP.}
\end{subfigure}%
\begin{subfigure}{.2\textwidth}
\centering
\captionsetup{justification=centering}
  \includegraphics[width=\linewidth]{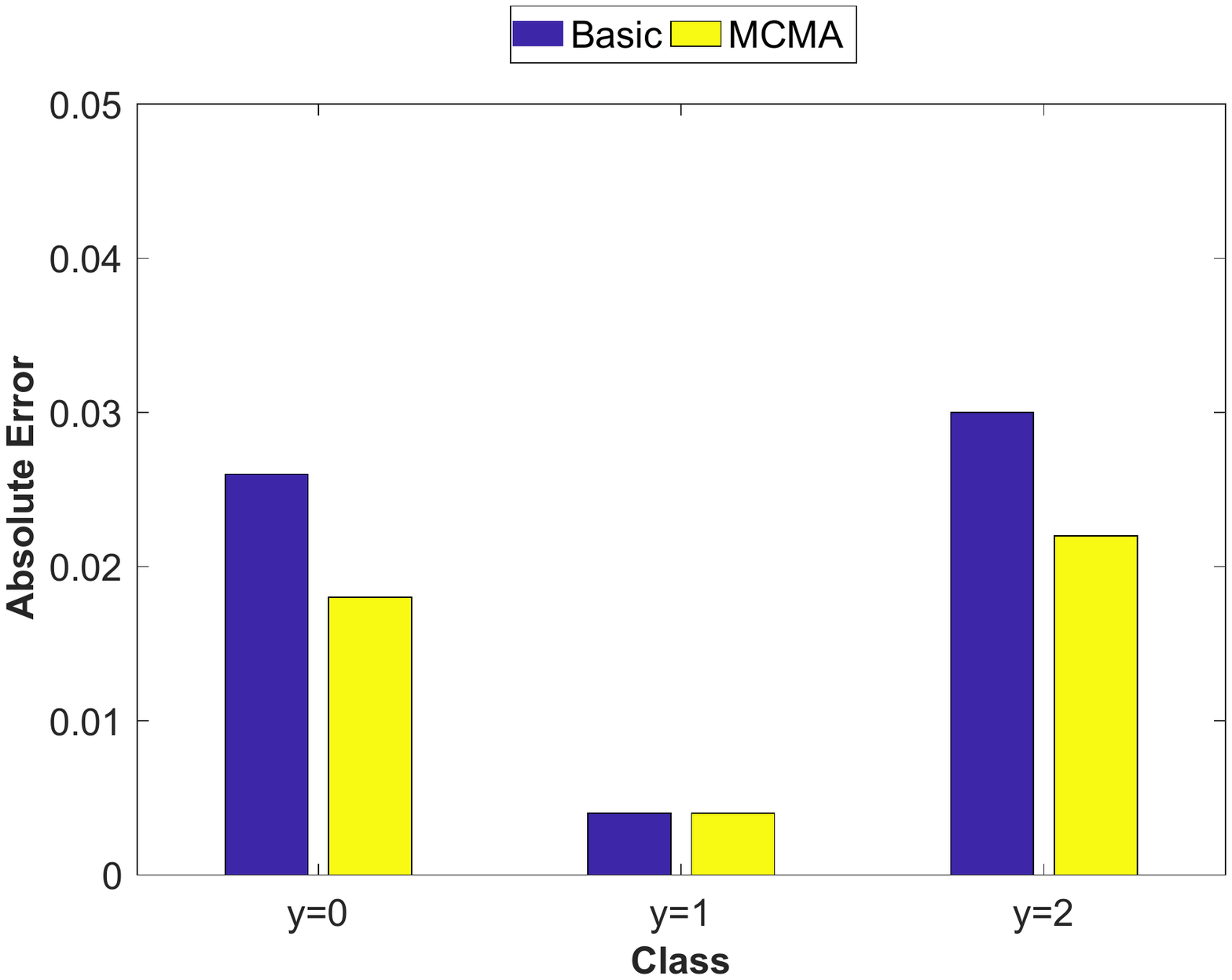}
    \caption{Gaussian NB.}
\end{subfigure}%
\begin{subfigure}{.2\textwidth}
\centering
\captionsetup{justification=centering}
  \includegraphics[width=\linewidth]{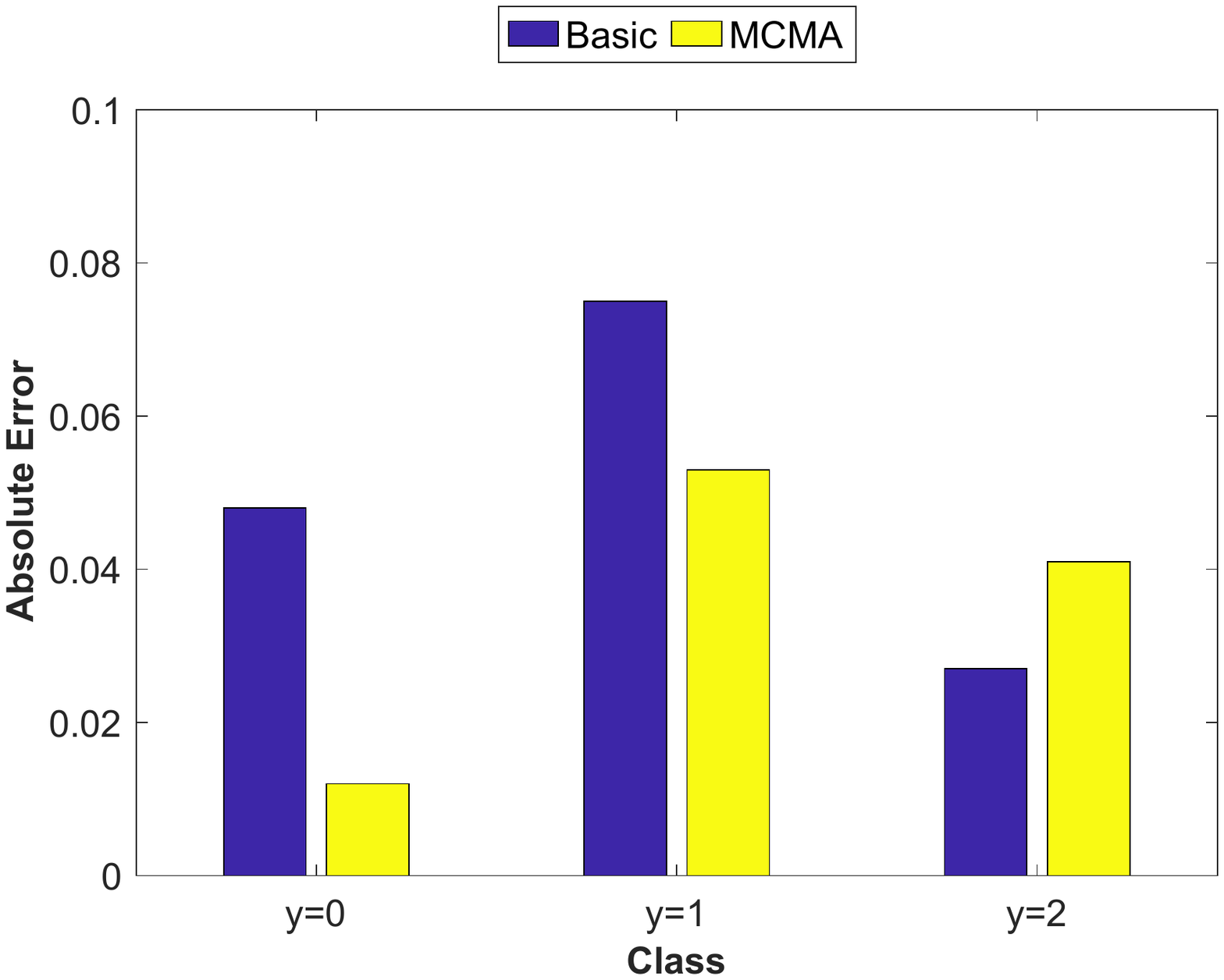}
    \caption{XGBoost.}
\end{subfigure}
\caption{Absolute Errors of the estimated probabilities of being the summary therapeutic association with $N=1000$.}
\label{1000}
\end{figure*}

\noindent\textbf{Predicting the \textit{Summary} Therapeutic Association (P3).}
The goal of meta-analysis is inferring the summary therapeutic association between the treatment/risk factor and the disease of interest. With the known data generating process, we can thus compare the estimated probabilities of each class being the summary therapeutic association $\hat{p}^s_k\ (k\in\{1,2,3\})$ with the ground-truth probabilities when $\bm{a}=\textbf{0}$. In the real world, the number of available RCTs varies. Thus, we evaluate across sample sizes $N\in \{100,500,1000\}$. Absolute error is used to measure the discrepancies: 
\begin{equation}
    \text{Absolute Error}=|p^s_k-\hat{p}^s_k| \quad k\in\{1,2,3\}.
\end{equation}
The results can be seen in Fig.\ \ref{100}-\ref{1000}. 

We begin by observing that the embedded causal perspective confers an advantage to using the substitute confounder, cf., the results for XGBoost with $N=500$. The best performance w.r.t. each probability prediction is achieved by various models whereas MCMA mostly presents at least two lowest estimation errors among the three classes ($p^s_1,p^s_2,p^s_3$), e.g., MCMA based on MNLogit, $k$-NN, and Gaussian NB with $N=500,1000$. The improvement is more significant in terms of $Y=0$ (i.e., the ground-truth summary therapeutic association) when the number of sample size $N$ is larger. These results suggest that (1) it is challenging to simultaneously optimize probabilities prediction for all three classes; (2) the overall performance improvement of MCMA over Basic multi-class classifiers corroborates the effectiveness of the embedded causal mechanisms in MCMA; and (3) in the era of big data, MCMA presents great potentials to accurately predict the summary therapeutic associations. 
\subsection{Semi-Synthetic Experiments (P4)}
Due to the difficulty of empirical evaluation using real-world data, here, we introduce an alternative using semi-synthetic data generated from real-world meta-analyses. To get the ground truth for the summary therapeutic association, we collect the scientific papers of meta-analysis for specific treatment/risk factor-disease pairs from PubMed 
and Cochrane Library databases. In doing so, we assume that all the published papers are selected based on strict reviewing processes, the results thereby should be in high quality. The ground truth for summary therapeutic associations is obtained from these meta-analysis papers. The next step identifies and collects the primary studies included in each meta-analysis. The RobotReviewer then takes these primary studies as input and outputs the risks of biases matrix $\mat{A}$ and therapeutic association $Y$. The last step of MCMA predicts the summary therapeutic association using the fitted outcome model in the Deconfounder. 

The number of existing meta-analysis papers consisting exclusively of RCTs is limited. We eventually managed to collect 54 such meta-analysis papers. Preliminary data analysis shows that the average number of RCTs included in each meta-analysis is 13, which is insufficient to train a classification model. To augment the data, we first gather the statistics obtained from these real-world RCTs and then use them to parameterize the simulation process. For example, the meta-analysis of PDE5 inhibitor and cardiac morphology in \citep{giannetta2014chronic} includes 18 RCTs, based on which the authors suggest positive summary therapeutic association between PDE5 inhibitor and cardiac morphology, i.e., $Y=2$. The simulation process starts with the risks of bias and therapeutic association of these 18 RCTs extracted from the RobotReviewer. We parameterize the Bernoulli distribution and the Multinomial distribution with the estimates from the real-world data to generate risks of bias $\mat{A}$ and therapeutic association $Y$, respectively. The sample size in this experiment is set to 100. 

With unknown probabilities of each category being the summary therapeutic association, we evaluate the models' performance on predicting therapeutic association in the RCTs and the summary therapeutic association. Results in Table \ref{PDE5} suggest findings that are similar to those we get from synthetic data. For instance, MCMA does not hurt the performance of but often outperforms the basic classification models on both AUC and F1 scores.
\vspace{-1mm}
\section{Discussion}
The empirical results shown in this work demonstrate the efficacy of using causal models to process the outputs of NLP-based data extraction and achieve the goal of meta-analysis: estimating \textit{summary therapeutic association} in the presence of hidden confounding bias. While there are other risks of bias not captured by existing NLP systems, currently identified biases can provide supporting evidence for existing conjectures and explore new sources of biases for further investigation. 

Causality-guided multi-class classifiers do not underperform basic classifiers in predicting individual therapeutic association and often outperform them when predicting summary therapeutic associations. It suggests that via the lens of multiple causal inference, we have a better understanding of the underlying data generating process involving risks of bias and therapeutic association. In contrast to correlation, causation implies that bias and therapeutic association have a cause-and-effect relationship with one another. Furthermore, controlling for hidden confounders presents an opportunity to alleviate the influence of both the ``known unknown'' (known but unmeasured confounders) and ``unknown unknown'' (unknown confounders) in the process of automating meta-analysis. Incorporating these factors will ensure more reliable and consistent prediction of summary therapeutic association, as suggested by both synthetic and semi-synthetic experiments.

The results here are not without limitations. As with other causal learning models relying on assumptions, the single-ignorability assumption in MCMA may be violated in practice. There might be hidden confounders that only influence one domain of risk of bias. Furthermore, limited by the functionality of RobotReviewer, the current implementation focuses solely on RCTs, rendering numerous observational studies unexplored. While the meta-analysis results in scientific publications are considered high-quality, there are inherent issues of publication and selection biases \citep{carter2019deep}. Moreover, MCMA relies on meta-analysts for literature search and paper relevance identification. Therefore, tools for automatic paper collection given the treatment and outcome are still needed. 

Nevertheless, this work advances collaborative research efforts in automating meta-analysis, which is currently a manual and complicated process. In this study, we provide initial solutions, but much work remains to be done. With our work, we can potentially discover effective and personalized therapies, improve the timeliness of clinical guidelines and even develop new directions for medical research and other domains of social good. We hope to bring concerns to the fore and broaden  discussions about the potential research directions of automated meta-analysis.
 \section*{Acknowledgements}
This work was conducted under the auspices of the IBM Science for Social Good initiative. The authors thank Iain Marshall and Byron Wallace for help with RobotReviewer and the staff of Reboot Rx for discussions.
\bibliography{uai2021-template}

\begin{thebibliography}{38}
\providecommand{\natexlab}[1]{#1}
\providecommand{\url}[1]{\texttt{#1}}
\expandafter\ifx\csname urlstyle\endcsname\relax
  \providecommand{\doi}[1]{doi: #1}\else
  \providecommand{\doi}{doi: \begingroup \urlstyle{rm}\Url}\fi

\bibitem[Abadi et~al.(2016)Abadi, Agarwal, Barham, Brevdo, Chen, Citro,
  Corrado, Davis, Dean, Devin, et~al.]{abadi2016tensorflow}
Mart{\'\i}n Abadi, Ashish Agarwal, Paul Barham, Eugene Brevdo, Zhifeng Chen,
  Craig Citro, Greg~S Corrado, Andy Davis, Jeffrey Dean, Matthieu Devin, et~al.
\newblock Tensorflow: Large-scale machine learning on heterogeneous distributed
  systems.
\newblock \emph{arXiv preprint arXiv:1603.04467}, 2016.

\bibitem[Athey et~al.(2019)Athey, Imbens, and Pollmann]{athey2019comment}
Susan Athey, Guido~W Imbens, and Michael Pollmann.
\newblock Comment on: ``{T}he {B}lessings of {M}ultiple {C}auses'' by {Yixin
  Wang and David M. Blei}.
\newblock \emph{Journal of the American Statistical Association}, 114\penalty0
  (528):\penalty0 1602--1604, 2019.

\bibitem[Borenstein et~al.(2011)Borenstein, Hedges, Higgins, and
  Rothstein]{borenstein2011introduction}
Michael Borenstein, Larry~V Hedges, Julian~PT Higgins, and Hannah~R Rothstein.
\newblock \emph{Introduction to meta-analysis}.
\newblock John Wiley \& Sons, 2011.

\bibitem[Carter(2019)]{carter2019deep}
Evan Carter.
\newblock Deep learning for robust meta-analytic estimation.
\newblock 2019.

\bibitem[Chen and Guestrin(2016)]{chen2016xgboost}
Tianqi Chen and Carlos Guestrin.
\newblock Xgboost: A scalable tree boosting system.
\newblock In \emph{KDD}, pages 785--794, 2016.

\bibitem[Dehejia and Wahba(2002)]{dehejia2002propensity}
Rajeev~H Dehejia and Sadek Wahba.
\newblock Propensity score-matching methods for nonexperimental causal studies.
\newblock \emph{Review of Economics and statistics}, 84\penalty0 (1):\penalty0
  151--161, 2002.

\bibitem[Giannetta et~al.(2014)Giannetta, Feola, Gianfrilli, Pofi, Dall’Armi,
  Badagliacca, Barbagallo, Lenzi, and Isidori]{giannetta2014chronic}
Elisa Giannetta, Tiziana Feola, Daniele Gianfrilli, Riccardo Pofi, Valentina
  Dall’Armi, Roberto Badagliacca, Federica Barbagallo, Andrea Lenzi, and
  Andrea~M Isidori.
\newblock Is chronic inhibition of phosphodiesterase type 5 cardioprotective
  and safe? a meta-analysis of randomized controlled trials.
\newblock \emph{BMC medicine}, 12\penalty0 (1):\penalty0 185, 2014.

\bibitem[Guo et~al.(2020)Guo, Cheng, Li, Hahn, and Liu]{guo2020survey}
Ruocheng Guo, Lu~Cheng, Jundong Li, P~Richard Hahn, and Huan Liu.
\newblock A survey of learning causality with data: Problems and methods.
\newblock \emph{CSUR}, 53\penalty0 (4):\penalty0 1--37, 2020.

\bibitem[Gutman and Rubin(2015)]{gutman2015estimation}
Roee Gutman and Donald~B Rubin.
\newblock Estimation of causal effects of binary treatments in unconfounded
  studies.
\newblock \emph{Statistics in medicine}, 34\penalty0 (26):\penalty0 3381--3398,
  2015.

\bibitem[Haidich(2010)]{haidich2010meta}
Anna-Bettina Haidich.
\newblock Meta-analysis in medical research.
\newblock \emph{Hippokratia}, 14\penalty0 (Suppl 1):\penalty0 29, 2010.

\bibitem[Higgins et~al.(2011)Higgins, Altman, G{\o}tzsche, J{\"u}ni, Moher,
  Oxman, Savovi{\'c}, Schulz, Weeks, and Sterne]{higgins2011cochrane}
Julian~PT Higgins, Douglas~G Altman, Peter~C G{\o}tzsche, Peter J{\"u}ni, David
  Moher, Andrew~D Oxman, Jelena Savovi{\'c}, Kenneth~F Schulz, Laura Weeks, and
  Jonathan~AC Sterne.
\newblock The cochrane collaboration’s tool for assessing risk of bias in
  randomised trials.
\newblock \emph{Bmj}, 343:\penalty0 d5928, 2011.

\bibitem[Hill(2011)]{hill2011bayesian}
Jennifer~L Hill.
\newblock Bayesian nonparametric modeling for causal inference.
\newblock \emph{Journal of Computational and Graphical Statistics}, 20\penalty0
  (1):\penalty0 217--240, 2011.

\bibitem[Holland(1986)]{holland1986statistics}
Paul~W Holland.
\newblock Statistics and causal inference.
\newblock \emph{JASA}, 81\penalty0 (396):\penalty0 945--960, 1986.

\bibitem[Imbens and Rubin(2015)]{imbens2015causal}
Guido~W Imbens and Donald~B Rubin.
\newblock \emph{Causal inference in statistics, social, and biomedical
  sciences}.
\newblock Cambridge University Press, 2015.

\bibitem[Kingma and Ba(2014)]{kingma2014adam}
Diederik~P Kingma and Jimmy Ba.
\newblock Adam: A method for stochastic optimization.
\newblock \emph{arXiv preprint arXiv:1412.6980}, 2014.

\bibitem[Lechner(2001)]{lechner2001identification}
Michael Lechner.
\newblock Identification and estimation of causal effects of multiple
  treatments under the conditional independence assumption.
\newblock In \emph{Econometric evaluation of labour market policies}, pages
  43--58. Springer, 2001.

\bibitem[Marshall et~al.(2016)Marshall, Kuiper, and
  Wallace]{marshall2016robotreviewer}
Iain~J Marshall, Jo{\"e}l Kuiper, and Byron~C Wallace.
\newblock Robotreviewer: evaluation of a system for automatically assessing
  bias in clinical trials.
\newblock \emph{JAMIA}, 23\penalty0 (1):\penalty0 193--201, 2016.

\bibitem[McCaffrey et~al.(2013)McCaffrey, Griffin, Almirall, Slaughter,
  Ramchand, and Burgette]{mccaffrey2013tutorial}
Daniel~F McCaffrey, Beth~Ann Griffin, Daniel Almirall, Mary~Ellen Slaughter,
  Rajeev Ramchand, and Lane~F Burgette.
\newblock A tutorial on propensity score estimation for multiple treatments
  using generalized boosted models.
\newblock \emph{Statistics in medicine}, 32\penalty0 (19):\penalty0 3388--3414,
  2013.

\bibitem[Michelson(2014)]{michelson2014automating}
Matthew Michelson.
\newblock Automating meta-analyses of randomized clinical trials: a first look.
\newblock In \emph{2014 AAAI Fall Symposium Series}, 2014.

\bibitem[Moher et~al.(2009)Moher, Liberati, Tetzlaff, Altman, Group,
  et~al.]{moher2009preferred}
David Moher, Alessandro Liberati, Jennifer Tetzlaff, Douglas~G Altman, Prisma
  Group, et~al.
\newblock Preferred reporting items for systematic reviews and meta-analyses:
  the prisma statement.
\newblock \emph{PLoS med}, 6\penalty0 (7):\penalty0 e1000097, 2009.

\bibitem[Ogburn et~al.(2019)Ogburn, Shpitser, and Tchetgen]{ogburn2019comment}
Elizabeth~L Ogburn, Ilya Shpitser, and Eric J~Tchetgen Tchetgen.
\newblock {Comment on ``The Blessings of Multiple Causes''}.
\newblock \emph{Journal of the American Statistical Association}, 114\penalty0
  (528):\penalty0 1611--1615, 2019.

\bibitem[Pearl(2019)]{pearl2019seven}
Judea Pearl.
\newblock The seven tools of causal inference, with reflections on machine
  learning.
\newblock \emph{Communications of the ACM}, 62\penalty0 (3):\penalty0 54--60,
  2019.

\bibitem[Pearl et~al.(2009)]{pearl2009causal}
Judea Pearl et~al.
\newblock Causal inference in statistics: An overview.
\newblock \emph{Statistics surveys}, 3:\penalty0 96--146, 2009.

\bibitem[Pildal et~al.(2007)Pildal, Hrobjartsson, J{\o}rgensen, Hilden, Altman,
  and G{\o}tzsche]{pildal2007impact}
J~Pildal, Asbj{\o}rn Hrobjartsson, KJ~J{\o}rgensen, J{\o}rgen Hilden,
  DG~Altman, and PC~G{\o}tzsche.
\newblock Impact of allocation concealment on conclusions drawn from
  meta-analyses of randomized trials.
\newblock \emph{Int J Epidemiol}, 36\penalty0 (4):\penalty0 847--857, 2007.

\bibitem[Ranganath and Perotte(2018)]{ranganath2018multiple}
Rajesh Ranganath and Adler Perotte.
\newblock Multiple causal inference with latent confounding.
\newblock \emph{arXiv preprint arXiv:1805.08273}, 2018.

\bibitem[Rassen et~al.(2011)Rassen, Solomon, Glynn, and
  Schneeweiss]{rassen2011simultaneously}
Jeremy~A Rassen, Daniel~H Solomon, Robert~J Glynn, and Sebastian Schneeweiss.
\newblock Simultaneously assessing intended and unintended treatment effects of
  multiple treatment options: a pragmatic “matrix design”.
\newblock \emph{Pharmacoepidemiology and drug safety}, 20\penalty0
  (7):\penalty0 675--683, 2011.

\bibitem[Rosenbaum and Rubin(1984)]{rosenbaum1984reducing}
Paul~R Rosenbaum and Donald~B Rubin.
\newblock Reducing bias in observational studies using subclassification on the
  propensity score.
\newblock \emph{JASA}, 79\penalty0 (387):\penalty0 516--524, 1984.

\bibitem[Rubin(1980)]{rubin1980randomization}
Donald~B Rubin.
\newblock Randomization analysis of experimental data: The fisher randomization
  test comment.
\newblock \emph{JASA}, 75\penalty0 (371):\penalty0 591--593, 1980.

\bibitem[Rubin(1990)]{rubin1990comment}
Donald~B Rubin.
\newblock Comment: Neyman (1923) and causal inference in experiments and
  observational studies.
\newblock \emph{Statistical Science}, 5\penalty0 (4):\penalty0 472--480, 1990.

\bibitem[Sackett et~al.(1996)Sackett, Rosenberg, Gray, Haynes, and
  Richardson]{sackett1996evidence}
David~L Sackett, William~MC Rosenberg, JA~Muir Gray, R~Brian Haynes, and
  W~Scott Richardson.
\newblock Evidence based medicine: what it is and what it isn't, 1996.

\bibitem[Seabold and Perktold(2010)]{seabold2010statsmodels}
Skipper Seabold and Josef Perktold.
\newblock statsmodels: Econometric and statistical modeling with python.
\newblock In \emph{9th Python in Science Conference}, 2010.

\bibitem[Song et~al.(2015)Song, Hao, and Storey]{song2015testing}
Minsun Song, Wei Hao, and John~D Storey.
\newblock Testing for genetic associations in arbitrarily structured
  populations.
\newblock \emph{Nature genetics}, 47\penalty0 (5):\penalty0 550--554, 2015.

\bibitem[Tipping and Bishop(1999)]{tipping1999probabilistic}
Michael~E Tipping and Christopher~M Bishop.
\newblock Probabilistic principal component analysis.
\newblock \emph{Statistical Methodology}, 61\penalty0 (3):\penalty0 611--622,
  1999.

\bibitem[Tran and Blei(2017)]{tran2017implicit}
Dustin Tran and David~M Blei.
\newblock Implicit causal models for genome-wide association studies.
\newblock \emph{arXiv preprint arXiv:1710.10742}, 2017.

\bibitem[Wang and Blei(2019)]{wang2019blessings}
Yixin Wang and David~M Blei.
\newblock The blessings of multiple causes.
\newblock \emph{JASA}, 114\penalty0 (528):\penalty0 1574--1596, 2019.

\bibitem[Xiong et~al.(2018)Xiong, Liu, Tse, Gong, Gladding, Smaill, Stiles,
  Gillis, and Zhao]{xiong2018machine}
Zhaohan Xiong, Tong Liu, Gary Tse, Mengqi Gong, Patrick~A Gladding, Bruce~H
  Smaill, Martin~K Stiles, Anne~M Gillis, and Jichao Zhao.
\newblock A machine learning aided systematic review and meta-analysis of the
  relative risk of atrial fibrillation in patients with diabetes mellitus.
\newblock \emph{Frontiers in physiology}, 9:\penalty0 835, 2018.

\bibitem[Yu et~al.(2006)Yu, Pressoir, Briggs, Bi, Yamasaki, Doebley, McMullen,
  Gaut, Nielsen, Holland, et~al.]{yu2006unified}
Jianming Yu, Gael Pressoir, William~H Briggs, Irie~Vroh Bi, Masanori Yamasaki,
  John~F Doebley, Michael~D McMullen, Brandon~S Gaut, Dahlia~M Nielsen, James~B
  Holland, et~al.
\newblock A unified mixed-model method for association mapping that accounts
  for multiple levels of relatedness.
\newblock \emph{Nature genetics}, 38\penalty0 (2):\penalty0 203--208, 2006.

\bibitem[Zanutto et~al.(2005)Zanutto, Lu, and Hornik]{zanutto2005using}
Elaine Zanutto, Bo~Lu, and Robert Hornik.
\newblock Using propensity score subclassification for multiple treatment doses
  to evaluate a national antidrug media campaign.
\newblock \emph{JEBS}, 30\penalty0 (1):\penalty0 59--73, 2005.

\end{thebibliography}
\clearpage
\appendix
\section{Justification of Deconfounder in Automated Meta-Analysis}
According to \citep{ogburn2019comment}, Deconfounder is justified to use under several conditions/assumptions in addition to the assumptions explicitly described in the original paper \citep{wang2019blessings}. In this appendix, we review these conditions/assumptions and show that they are satisfied in our problem setting.
\begin{enumerate}
    \item \textit{No M-bias collider exists between A and Y.}
    \item \textit{No post-treatment variables are captured by Z.}
 \item \textit{The causes are not causally dependent.}
    \item \textit{Unmeasured multi-cause confounding is due to a dependence or clustering structure that is common
to each cause and to the outcome.}
    \item \textit{Z is consistent.}
    \item \textit{In the limit as the number of causes and the number of observations go to infinity.}
\end{enumerate}

Assumptions 1-2 requires that all covariates used in Deconfounder are pre-treatment variables and no post-treatment variables shall exist. Different from conventional data used in causal inference, the causes and outcome in our problem are simply the risks of bias and therapeutic associations \textit{simultaneously} extracted from scientific studies by NLP systems. Existence of post-treatment variables are out of questions. 

Assumption 3 is satisfied because the pre-defined risks of bias are causally independent by definition. 

For Assumption 4, we consider the major unmeasured multi-cause confounding in automated meta-analysis from the preconceived bias due to domain expertise and the uncertainty of the NLP systems. Both are common to each cause and to the outcome. 

Assumptions 5-6 require consistency of substitute confounders, which generally holds asymptotically as the number of risks of bias goes to infinity. 

Despite some assumptions being untestable in practice, as the first work in automated meta-analysis with extremely limited information, this study aims to bring to light the possibility of applying causal learning to achieving the goal of automated meta-analysis.
\end{document}